\documentclass[twocolumn,showpacs]{revtex4-1}

\usepackage{color}
\usepackage{graphicx}
\usepackage{epstopdf}
\usepackage{dcolumn}
\usepackage{amsmath}
\usepackage[caption=false]{subfig}

\begin{document}
\bibliographystyle{revtex}


\title[Short Title]{Phase transitions and symmetry energy in nuclear pasta}

\author{C.O. Dorso}
\affiliation{Instituto de F\'\i sica de Buenos Aires, Pabell\'on
I, Ciudad Universitaria, 1428 Buenos Aires, Argentina. }

\author{G.A. Frank}
\affiliation{Unidad de Investigaci\'on y Desarrollo de las
Ingenier\'\i as, Universidad Tecnol\'ogica Nacional, Facultad Regional Buenos
Aires, Av. Medrano 951, 1179 Buenos Aires, Argentina. }

\author{J.A. L\'opez}
\affiliation{University of Texas at El Paso, El Paso, Texas 79968, U.S.A.}

\date{\today}

\begin{abstract}

Cold and isospin-symmetric nuclear matter at sub-saturation densities is known 
to form the so-called pasta structures, which, in turn, are known to undergo 
peculiar phase transitions. Here we investigate if such pastas and their phase 
changes survive in isospin asymmetric nuclear matter, and whether the symmetry 
energy of such pasta configurations is connected to the isospin content, the 
morphology of the pasta and to the phase transitions. We find that indeed pastas 
are formed in isospin asymmetric systems with proton to neutron ratios of 
$x=0.3$, 0.4 and 0.5, densities in the range of 0.05 fm${}^{-3}<\rho<$ 0.08 
fm${}^{-3}$, and temperatures $T<$ 2 MeV. Using tools (such as the caloric 
curve, Lindemann coefficient, radial distribution function, Kolmogorov 
statistic, and Euler functional) on the composition of the pasta, determined the 
existence of homogeneous structures, tunnels, empty regions, cavities and 
transitions among these regions. The symmetry energy was observed to attain 
different values in the different phases showing its dependence on the 
morphology of the nuclear matter structure. 

\end{abstract}

\maketitle

\section{Introduction}\label{intro}

The effect of the excess of neutrons to protons in the nuclear equation of
state (EOS) is characterized by the symmetry energy,
$E_{sym}(T,\rho)$, and its importance in phenomena ranging from
nuclear structure to astrophysical processes has prompted intense
investigations~\cite{li,EPJA}. Some of the latest experimental and
theoretical studies of the symmetry energy have been at subsaturation
densities and warm temperatures~\cite{natowitz,lopez2017}; the behavior
of the symmetry energy at even lower temperatures is still unknown
and it is the subject of the present investigation.\\

Nuclear systems exhibit fascinating complex phenomena at subsaturation
densities and warm and cold temperatures. At densities below the saturation
density, $\rho_0 = 0.16\,\mathrm{fm}^{-3}$, and temperatures, say, between
1 MeV and 5 MeV, nuclear systems are well inside a condensed region 
and can undergo phase transitions.\\

Experimental reactions~\cite{kowa,wada,natowitz} have shown that $E_{sym}$ is
affected by the formation of clusters. A recent calculation of the symmetry
energy at clustering densities and temperatures~\cite{lopez2017} obtained
good agreement with experimental data, corroborating the Natowitz
conjecture~\cite{natowitz,nato2015}, namely that the asymptotic limit of
$E_{sym}$ would not tend to zero at small densities as predicted by mean-field
theories.\\

The problem of estimating the symmetry energy at even lower temperatures is
even more challenging. At colder temperatures ($T<1\,$MeV) nuclear systems
are theorized to form the so-called ``nuclear pasta'', which are of interest
in the physics of neutron stars~\cite{ravenhall}. Since neutron star
cooling is due mostly to neutrino emission from the core, the interaction
between neutrinos and the crust pasta structure is bound to be relevant
in the thermal evolution of these stars~\cite{dorso2017}. An additional
challenge of the study of $E_{sym}$ in these cold and sparse systems is that
nucleons in the pasta have been found to undergo phase transitions
between solid and liquid phases~\cite{dorso2014} {\it within the pasta
structures}.\\

To study the symmetry energy of such complex systems one must use models capable
of exhibiting particle-particle correlations that will lead to clustering
phenomena and phase changes. Even though most of the studies of $E_{sym}$
have been based on mean-field approaches~\cite{li}, the low temperature-low density investigations lie
outside the scope of these models as they fail to describe clustering phenomena.
To correct for this, some calculations have attempted to include a limited number
of cluster species by hand~\cite{hempel,horo-s,hempel2015}, by using thermal
models~\cite{agrawal}, or hybrid interpolations between methods with embedded cluster
correlations and mean field theories~\cite{natowitz}.\\

On the opposite side of the theoretical spectrum, the Classical Molecular Dynamics
(CMD) model is able to mimic nuclear systems and yield cluster formation without
any adjustable parameters. This is the model used, for instance, to calculated
$E_{sym}$ in the liquid-gas region~\cite{lopez2017}, and to find the
solid-liquid phase transition in the pasta~\cite{dorso2014}.\\

Thus, the questions that occupy us in the present investigation are: do the
different phases of the pasta structures found in
Ref.~\cite{dorso2014} survive in non-symmetric nuclear matter? And, how does the
symmetry energy behaves at low densities and temperatures? \textit{i.e.} within
the pasta. In this work we extend the low density calculations of the
symmetry energy of Ref.~\cite{lopez2017} to lower temperatures, and connect
them to the morphologic and thermodynamic properties of the pasta
found in Ref.~\cite{dorso2014}, but now at proton fractions in the range of
30$\%$ to 50$\%$.\\

\section{Nuclear pasta}\label{NP}

The nuclear pasta refers to structures produced by the spatial
arrangement of protons and neutrons, theorized to exist in neutron star
crusts. These structures form when the nucleon configuration reaches a
free energy minima, and can be calculated by finding the minimum of the energy
using static methods (usually at zero temperature), or by a cooling bath using
dynamical numerical methods. Although a pasta can be formed in pure
nuclear matter (\textit{i.e.} only protons and neutrons) due to the competition
between the repulsive and attractive nuclear
forces~\cite{lopez2015,gimenez2014}, the term nuclear pasta usually
refers to the structures expected to exist in neutron star matter (NSM)
composed of nuclear matter embedded in an electron gas.\\

The nuclear pasta was first predicted through the use of the liquid
drop model~\cite{ravenhall,Hashimoto}, which helped to find the most
common ``traditional'' pasta structures, such as \textit{lasagna},
\textit{spaghetti} and \textit{gnocchi}. These findings were corroborated with
mean field~\cite{Page} and Thomas-Fermi models~\cite{Koonin}, and later
with quantum molecular dynamics~\cite{Maruyama,Kido,Wata2002} and classical
potential models~\cite{Horo2004,dor12, dor12A}. More recent applications of the
classical molecular dynamics (CMD) allowed the detection of different
``non-traditional'' pasta phases~\cite{dorso2014}.\\

As found in Ref.~\cite{dorso2014}, dynamic models (such as CMD) can
drive the neutron star matter into free energy minima, which abound in complex
energy landscapes. Local minima are usually surrounded by energy barriers which
effectively trap cold systems. It must be remarked that most static models are
unable to identify these local minima as actual equilibrium solutions, while
dynamical models can identify these local minima by careful cooling. Thus, as
proposed in Ref.~\cite{dorso2014}, at low but finite temperatures the state of
the system should be described as an ensemble of both traditional and
nontraditional (amorphous, sponge-like) structures rather than by a single one.

\subsection{Simulating the nuclear matter}\label{cmd}

The advantages of classical molecular dynamics (CMD) over other numerical and
non-numerical methods have been presented
elsewhere~\cite{dor12,dor12A,dorso2014}, as well as the validity of its
classical approach in the range of temperatures and densities achieved in
intermediate-energy nuclear reactions~\cite{lopez2014}.\\

In a nutshell, CMD treats nucleons as classical particles
interacting through pair potentials and predicts their dynamics by
solving their equations of motion numerically.  The method does not
contain adjustable parameters, and uses the Pandharipande (Medium)
potentials~\cite{pandha}, of which one is attractive between
neutrons and protons, and the other one is repulsive between equal nucleons,
respectively. The corresponding expressions read\\

\begin{equation}
\left\{\begin{array}{rcl}
        V_{np}(r) & = &
\displaystyle\frac{V_{r}}{r}e^{-\mu_{r}r}-\displaystyle\frac{V_{r}}{r_c}e^{-\mu_
{ r } r_ { c } } -\displaystyle\frac{V_ { a }}{r}
e^{-\mu_{a}r}+\displaystyle\frac{V_{a}}{r_{c}}e^{-\mu_{a}r_{c}}\\
       & & \\
       V_{nn}(r) & = &
\displaystyle\frac{V_{0}}{r}e^{-\mu_{0}r}-\displaystyle\frac{V_{0}}{r_{c}}e^{
-\mu
_{0}r_{c}}
       \end{array}\right.
\end{equation}

\noindent where $r_c$ is the cutoff radius after which the potentials
are set to zero. The corresponding parameter values are summarized in 
Table~\ref{table_parameter}. These parameters were set by Pandharipande to 
produce a cold nuclear matter saturation density of $\rho_0=0.16$ fm${}^{-3}$, a 
binding energy $E(\rho_0)=-16$ MeV/nucleon and a compressibility of about 
$250\,$MeV.\\

\begin{table}
{\begin{tabular}{l @{\hspace{9mm}}@{\hspace{9mm}} r
@{\hspace{6mm}}@{\hspace{6mm}} l}
\toprule
      Parameter &  \multicolumn{1}{l}{Value} & \multicolumn{1}{l}{Units}    
\\
\colrule
$V_r$ &  3088.118 & MeV \\
$V_a$ &  2666.647 & MeV\\
$V_0$ &  373. 118 & MeV\\
$\mu_r$ & 1.7468 & fm$^{-1}$ \\
$\mu_a$ & 1.6000 & fm$^{-1}$ \\
$\mu_0$ & 1.5000 & fm$^{-1}$ \\
$r_c$   & 5.4    & fm \\
\botrule
\end{tabular}
}
\caption{Parameter set for the CMD computations. The values correspond to the 
Pandharipande Medium Model. }
\label{table_parameter}
\end{table}

\begin{figure*}[!htbp]
\centering
\subfloat[$\rho=0.05$\label{fig:rho05}]{
\includegraphics[width=0.57\columnwidth]
{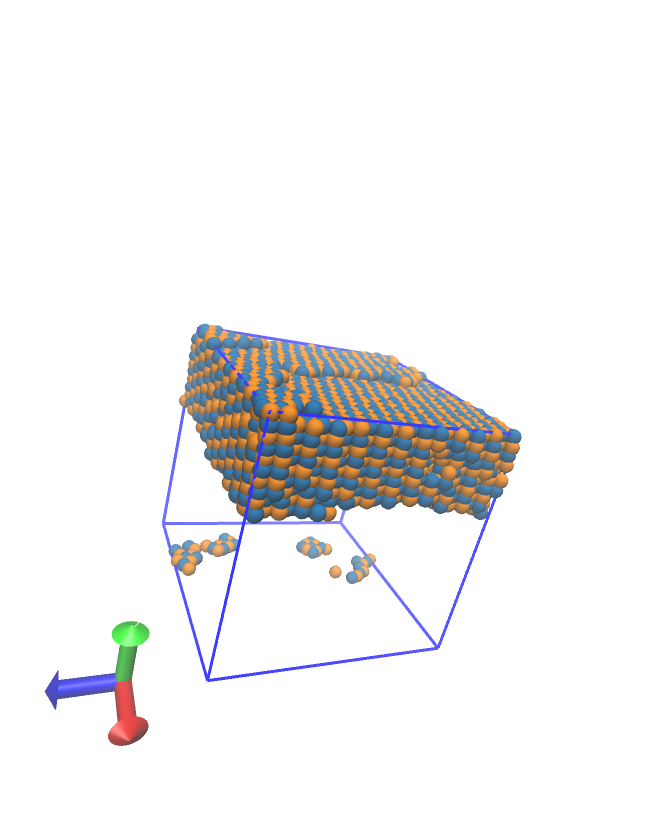}
}
\subfloat[$\rho=0.06$\label{fig:rho06}]{
\includegraphics[width=0.47\columnwidth]
{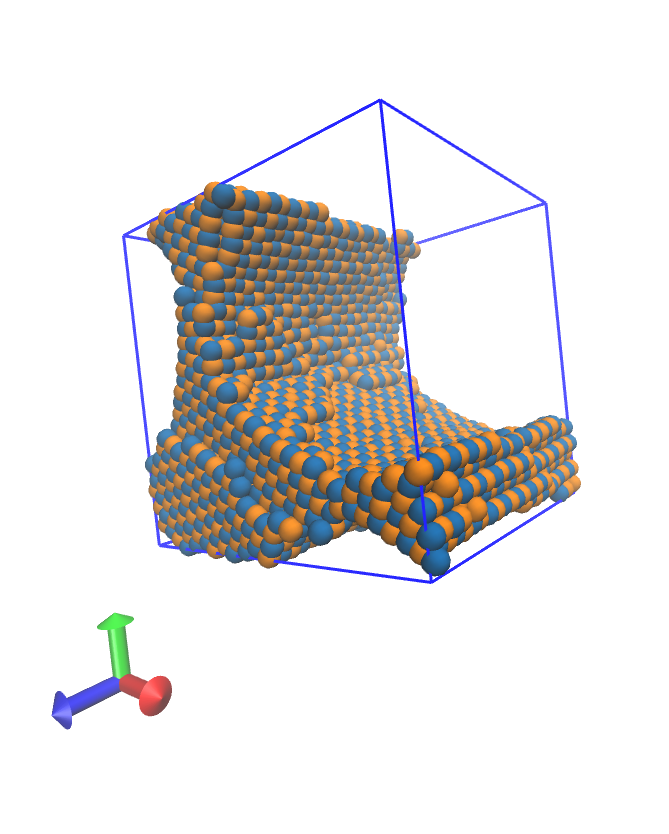}
}
\subfloat[$\rho=0.07$\label{fig:rho07}]{
\includegraphics[width=0.57\columnwidth]
{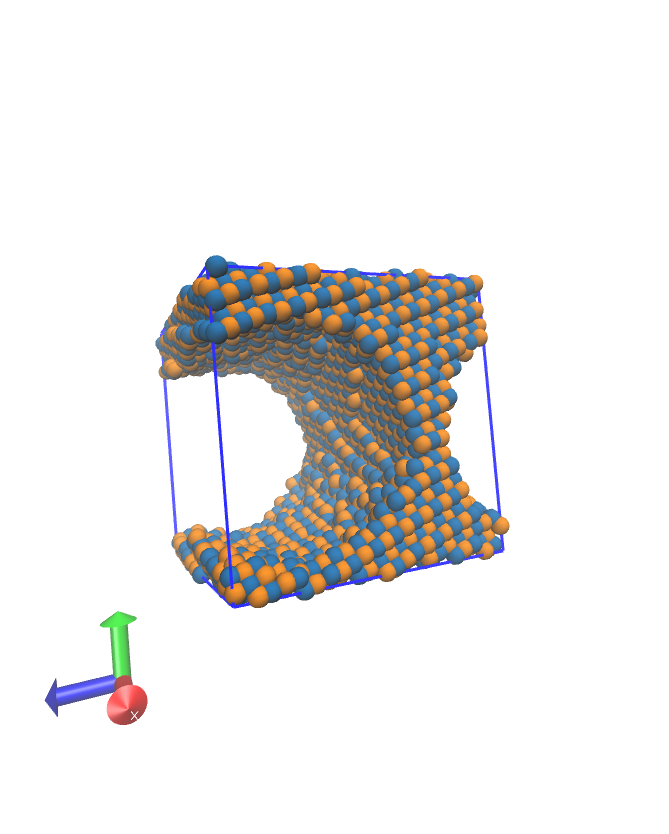}
}
\subfloat[$\rho=0.085$\label{fig:rho085}]{
\includegraphics[width=0.47\columnwidth]
{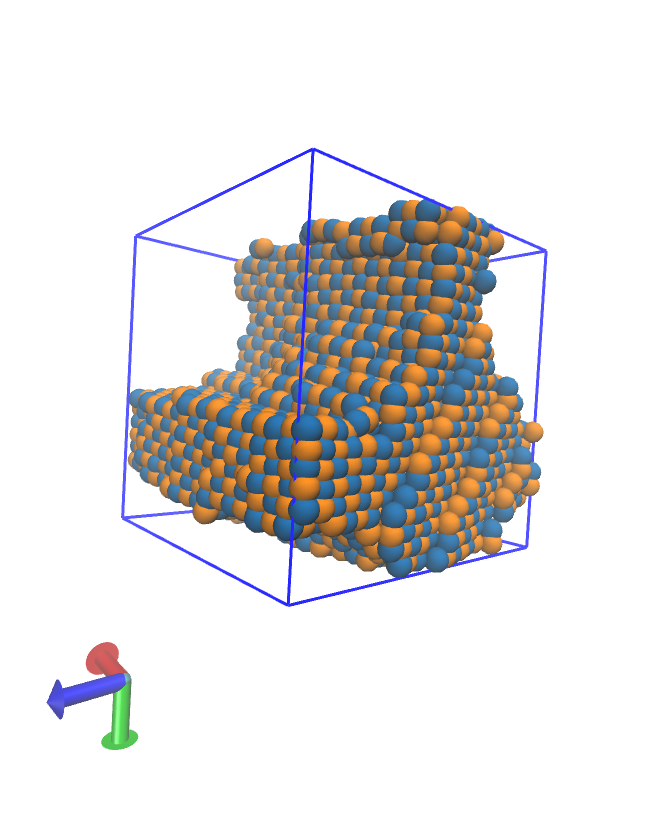}
}
\caption{\label{fig:pasta} (Color online) Pasta structures for
nuclear matter systems with 6000 nucleons,
$x=0.5$ at $T=0.2\,$MeV and densities $\rho$=0.05, 0.06, 0.07 and
0.085$\,$fm${}^{-3}$. Protons are represented in light color (orange), while 
neutrons are
represented in darker color (blue). The red arrow (out of the page) corresponds 
to the $x$ coordinate, the
green arrow (vertical) to the $y$ coordinate and the  blue one (horizontal) to 
the $z$ coordinate.}
\end{figure*}

Figure~\ref{fig:pasta} shows an example of the pasta structures for
nuclear matter with 6000 nucleons in the simulating cell (and periodic 
boundary conditions), $x=0.5$ at $T = 0.2$ MeV, and
densities $\rho$=0.05, 0.06, 0.07 and 0.085 fm${}^{-3}$, respectively.\\

We study the properties of a system of 6000 particles (with periodic 
boundary conditions) using the LAMMPS code~\cite{lammps} with the Pandharipande 
and screened Coulomb potentials. The total number of particles is divided into
protons (P) and neutrons (N) according to values of $x= P/(N+P) =$
0.3, 0.4 and 0.5. The nuclear system is cooled down from a
relatively high temperature ($T\geq 4.0\,$MeV) to a desired cool
temperature in small temperature steps ($\Delta T = 0.01\,$MeV) with the
Nos\'e Hoover thermostat~\cite{nose}, and assuring that the energy,
temperature, and their fluctuations are stable.\\

\begin{figure}
\begin{center}
   \includegraphics[width=3.5in]{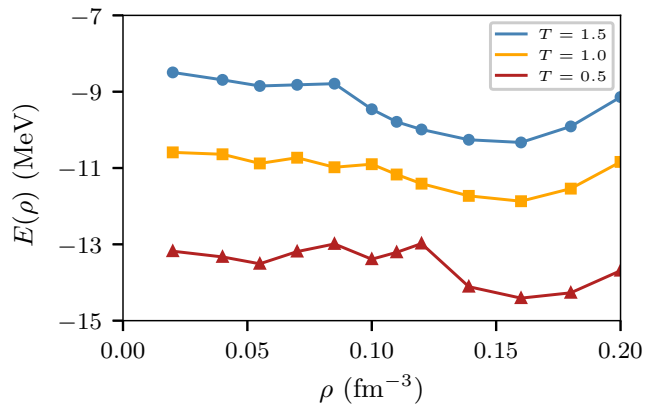}
\caption{Energy per nucleon versus density for nuclear matter with $x=0.5$ at
$T=1.5$ (circles), $1.0$ (squares) and $0.5$~MeV (triangles). The total number
of nucleons in the simulating cell was $N=2000$. The system was cooled from
2~MeV to 0.5~MeV.}\label{E-dens} 
\end{center}
\end{figure}

Figure~\ref{E-dens} shows an example of the energy per nucleon
versus the density for systems with 2000 particles at $x=0.5$ and $T =
1.5$, $1.0$ and $0.5\,$MeV. Clearly visible are the homogeneous phase 
(\textit{i.e.}
those under the ``$\cup$'' part of the energy-density curve), and the loss
of homogeneity at lower densities. \\

From Fig.~\ref{fig:pasta} and Fig.~\ref{E-dens} it is possible to distinguish 
three distinct regions to be analyzed. The first one, that goes from very low 
densities up to approximately $0.08\,$fm$^{-3}$ in which the system displays 
pasta structures, a crystal-like region at densities above approximately 
$0.14\,$fm$^{-3}$, and a transition region between these two~\cite{dor12A}.\\

\subsubsection{Comparison to neutron star matter}\label{cmd_star}

Although the present article focuses solely on nuclear matter, it is convenient 
to compare to neutron star matter, which includes a neutralizing electron gas 
embedding the nuclear matter.  The electron gas is included through a
screened Coulomb potential between protons (additional to the Pandharipande
potentials) of the Thomas-Fermi form~\cite{Horo2004,Maruyama,dor12}

\begin{equation}
V_{tf}=\displaystyle\frac{q^2}{r}e^{-r/\lambda}.
\end{equation}

\begin{figure*}[!htbp]
\centering
\subfloat[$\rho=0.05$\label{fig:rho05_star}]{
\includegraphics[width=0.5\columnwidth]
{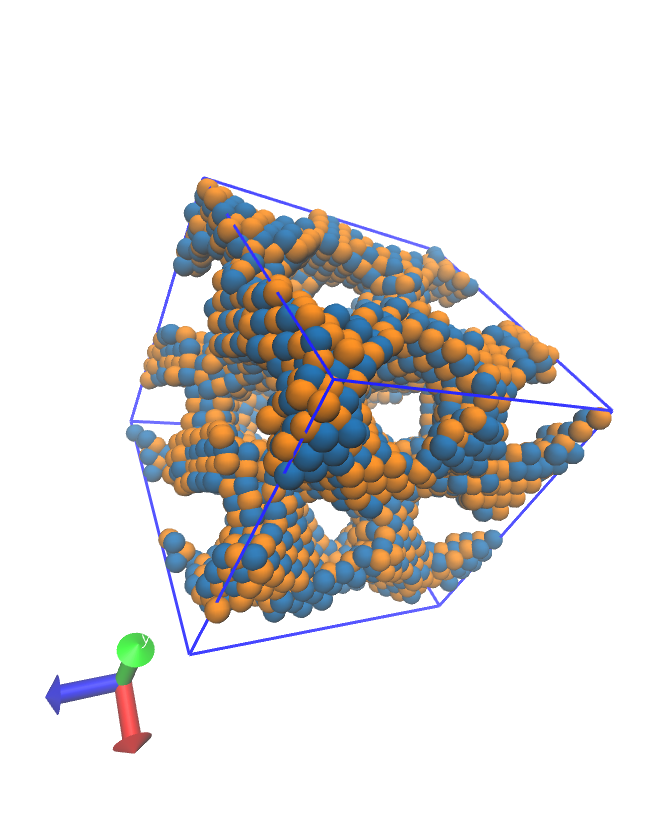}
} 
\subfloat[$\rho=0.06$\label{fig:rho06_star}]{
\includegraphics[width=0.5\columnwidth]
{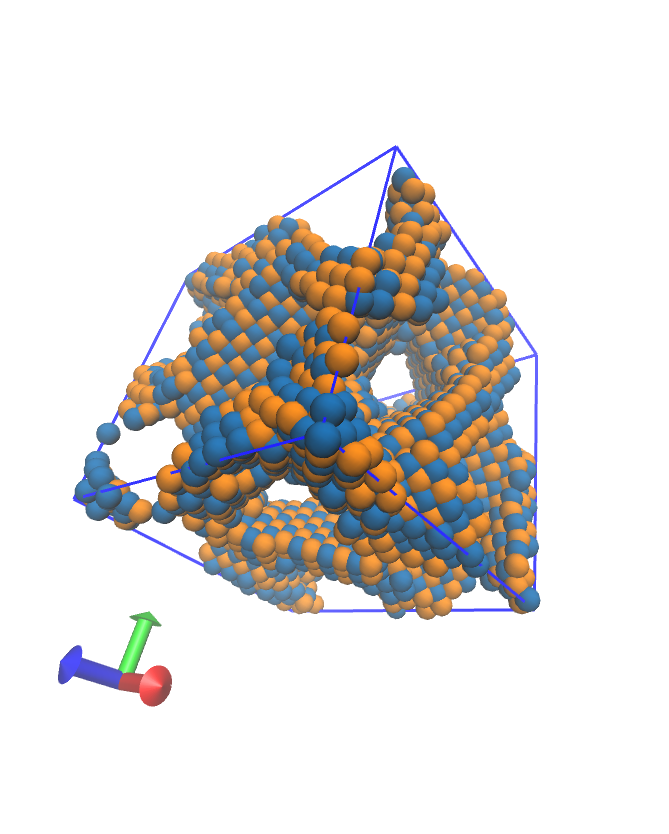}
}
\subfloat[$\rho=0.07$\label{fig:rho07_star}]{
\includegraphics[width=0.5\columnwidth]
{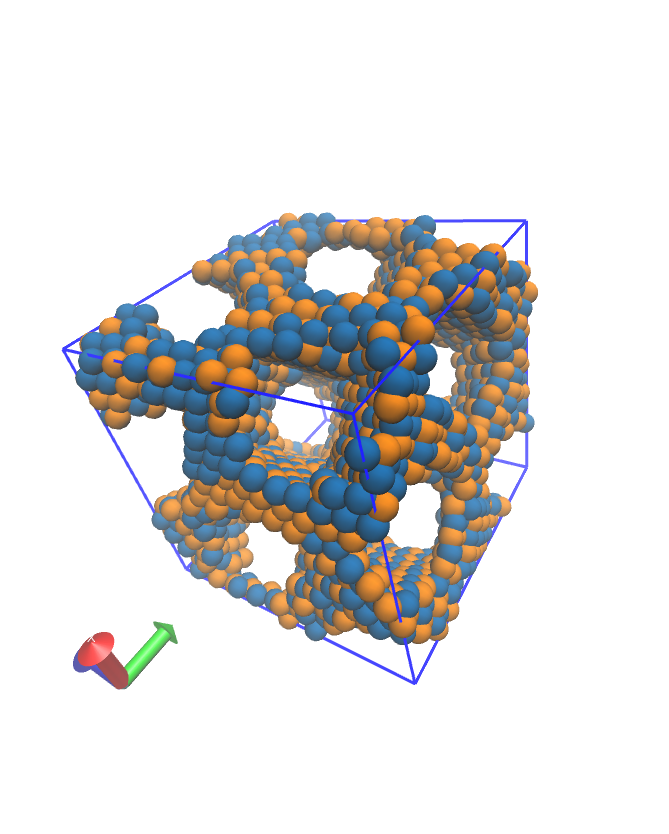}
}
\subfloat[$\rho=0.085$\label{fig:rho085_star}]{
\includegraphics[width=0.5\columnwidth]
{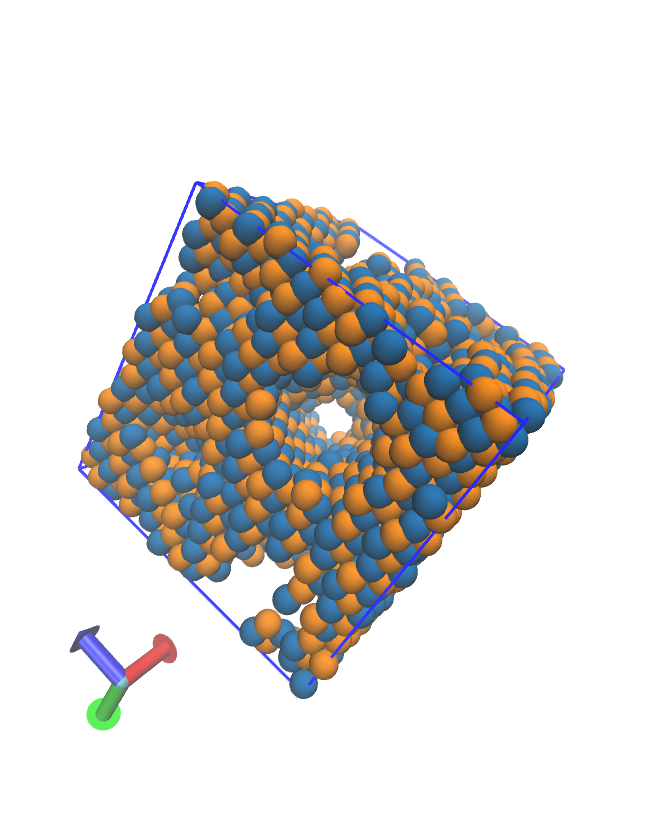}
}
\caption{\label{fig:pasta_star} (Color online) Pasta structures for
neutron star matter (with screened Coulomb potential) systems with 4000
nucleons, $x=0.5$ at $T=0.2\,$MeV and densities $\rho$=0.05, 0.06, 0.07 and
0.085$\,$fm${}^{-3}$. Protons are represented in orange, while neutrons are
represented in blue. The red arrow corresponds to the $x$ coordinate, the
green arrow to the $y$ coordinate and the  blue one to the $z$ coordinate. }
\end{figure*}

\noindent A screening length of $\lambda=20\,$fm is long
enough to reproduce the density fluctuations of this model~\cite{dor14},
likewise the cutoff distance for $V_{tf}$ of 20 fm. Figure~\ref{fig:pasta_star} 
shows an example of the pasta structures for symmetric neutron star matter 
at $T = 0.2$ MeV and densities $\rho$=0.05, 0.06, 0.07 and 0.085 
fm${}^{-3}$, respectively. A complete study of phase transitions and the 
symmetry energy in asymmetric neutron star crusts will presented in the near 
future.\\

\subsection{Characterization of the pasta regime}\label{tools}

To characterize the pasta, its phase changes and to calculate the
symmetry energy we use the the caloric curve, the radial distribution
function, the Lindemann coefficient, Kolmogorov statistic, Minkowski functionals
and a numerical fitting procedure to estimate $E_{sym}$. These techniques are
briefly reviewed next.

\subsubsection{Caloric Curve}\label{cal}

The heat capacity is a measure of the energy (heat) needed to increase
the temperature of the nuclear system, and the plot of energy versus
temperature is known as the caloric curve. The energy added to a system
is usually distributed among the degrees of freedom of all the particles
and, at some precise circumstances, such energy can be used to break
inter-particle bonds and make the system go from, say, a solid to a liquid
phase; during phase changes an increase of the energy does not result in an
increase of the temperature. Likewise, as in different phases there are
different number of degrees of freedom, the rate of heating (slope of the
caloric curve) is different in the different phases; changes of slope are
thus indicators of phase transitions. \\

\subsubsection{Radial distribution function}\label{pair}
The radial distribution function is the ratio of the local density of nucleons
to the average one, $g(r) = \rho (r)/\rho_0$, and it effectively describes the
average radial distribution of nucleons around each other. The computation
of $g(r)$ is as follows

\begin{equation}
 g(r)=\displaystyle\frac{V}{4\pi r^2N^2}\bigg\langle\displaystyle\sum_{i,
j}\delta(r-r_{ij})\bigg\rangle\label{eqn:gr_def}
\end{equation}

\noindent where $r_{ij}=|\mathbf{r}_{j}-\mathbf{r}_{i}|$ is the inter-nucleon
distance, while $V$ and $N$ are the simulation cell volume and the total number
of nucleons, respectively. Thus, $g(r)$ is obtained by tallying the distance
between neighboring nucleons at a given mean density and temperature, averaged
over a large number of particles. A rigorous definition of $g(r)$ for 
non-homogeneous systems, such as the pasta, can be found in 
Appendix~\ref{sec:gr_example}.  \\

The strength of the peaks of $g(r)$ help to detect whether nucleons are in a
regular (solid) phase or in a disordered (liquid) system. See
Section \ref{results_1} for details. \\

\subsubsection{Lindemann coefficient}\label{lind}
The Lindemann coefficient~\cite{lindemann} provides an estimation of
the root mean square displacement of the particles respect to their
equilibrium position in a crystal state, and it serves as an
indicator of the phase where the particles are in, as well as of transitions
from one phase to another. Formally,

\begin{equation}
\Delta_L =
\displaystyle\frac{1}{a}\displaystyle\sqrt{\displaystyle\sum_{i=1}
^N\bigg\langle\displaystyle\frac{\Delta r^2_i}{N}\bigg\rangle}
\end{equation}

\noindent where $\Delta r^2_i=(r_i-\langle r_i\rangle)^2$, $N$  is the number
of particles, and $a$ is the crystal lattice constant; for the nuclear case we
use the volume per particle to set the length scale through $a = (V/N)^{1/3}$.
\\

\subsubsection{Kolmogorov statistic}\label{kolm}
The Kolmogorov statistic measures the difference between a sampled (cumulative)
distribution $F_n$ and a theoretical distribution $F$. The statistic, as
defined by Kolmogorov \cite{birnbaum}, applies to univariate distributions (1D)
as follows

\begin{equation}
D_N = \mathrm{sup}_{\{x\}}|F_n(x) - F(x)|
\end{equation}

\noindent where ``sup'' means the supremum value of the argument along the $x$
domain, and $N$ is the total number of samples. This definition is proven to
represent univocally the greatest absolute discrepancy between both
distributions. \\

An extension of the Kolmogorov statistic to multivariate distributions,
however, is not straight forward and researchers moved in different directions
for introducing an achievable statistic \cite{gosset}. The Franceschini's
version seems to be ``well-behaved enough'' to ensure that the computed
supremum varies in the same way as the ``true'' supremum. It also appears to be
sufficiently distribution-free for practical purposes \cite{franceschini}. \\

The three dimensional (3D) Franceschini's extension of the Kolmogorov statistic
computes the supremum for the octants

\begin{eqnarray}
 (x<X_i,y<Y_i,z&<&Z_i),   \nonumber
\\ (x<X_i,y&<&Y_i,z>Z_i),  \nonumber
\\ ...\ (x&>&X_i,y>Y_i,z>Z_i)
\end{eqnarray}

\noindent for any sample $(X_i,Y_i,Z_i)$, $i$ denoting each of the $N$ 
particles, 
and chooses the supremum from this set of eight values. The method assumes 
that the variables $X_i$, $Y_i$ and $Z_i$ are not highly correlated. \\

In the nuclear case, the Kolmogorov 3D (that is, the Franceschini's version)
quantifies the discrepancy in the nucleons positions compared to the
homogeneous case. \\

It is worth mentioning that the reliability of the 3D Kolmogorov statistic has
been questioned in recent years \cite{babu}. The arguments, however, focus on
the correct confidence intervals when applying the 3D Kolmogorov statistic to
the null-hypothesis tests. Our investigation does not require computing these
intervals, and thus, the questionings are irrelevant to the matter.  \\

\subsubsection{Minkowski functionals}\label{mink}
Minkowski functionals~\cite{michielsen} are functions designed to
measure the size, shape and connectivity of spatial structures
formed by the nucleons. For three dimensional bodies these functions
are the volume, surface area, Euler characteristic ($\chi$), and
integral mean curvature (B). The Euler characteristic can be
interpreted as

\begin{equation}
\chi = \mathrm{isolated\ regions} + \mathrm{cavities} -
\mathrm{tunnels}\label{eq:chi}
\end{equation}

\noindent while $B$ is a measure of the curvature of the surface of a given
structure. In Ref.~\cite {dor12A} it was found that the pasta
structures can be classified according to Table~\ref{tab1}.\\

\begin{table}
{\begin{tabular}{c @{\hspace{4mm}}|@{\hspace{4mm}} c
@{\hspace{5mm}}|@{\hspace{5mm}} c @{\hspace{5mm}}|@{\hspace{5mm}} c}
\toprule
           &   B $<$ 0  & B $\sim$ 0  &   B $>$ 0 \\
\colrule
$\chi > 0$ & anti-gnocchi &  & Gnocchi\\
$\chi \sim 0$ & anti-gpaghetti  & lasagna & spaghetti\\
$\chi < 0$ & anti-jungle gym &  & jungle gym \\
\botrule
\end{tabular}
}
\caption{Integral mean curvature and Euler characteristic values for
pasta shapes. The ``anti'' prefix means the inverted situation between
occupied and empty regions. The ``jungle gym'' stands for a 3D rectangular
trellis.}
\label{tab1}
\end{table}

\subsubsection{Symmetry energy}\label{esymm}

The evaluation of the symmetry energy follows the procedure
introduced before~\cite{lopez2014,lopez2017}. The symmetry energy
is defined as

\begin{equation}
E_{sym}(\rho,T)=\displaystyle\frac{1}{2!}\,\displaystyle
\frac{\partial^2E(\rho,T,\alpha)}{\partial \alpha^2}\bigg|_{\alpha=0}
\end{equation}

\noindent with $\alpha=(N-Z)/(N+Z)=1-2x$. Using the CMD results of the internal
energy $E(\rho,T,x)$ it is possible to construct a continuous
function by fitting the values of $E(T,\rho,\alpha)$ for each $T$ and
$x$ with an expression of the type

\begin{equation}
E(T,\rho,\alpha)= \displaystyle\sum_{i=0}^3E_i(T,\alpha)\,\rho^i
\end{equation}

The $\alpha$ dependence of the coefficients $E_i(T,\alpha)$ can be extracted
from the CMD data calculated at various values of $\alpha$, and assuming an
$\alpha$ dependence of the type

\begin{equation}
E_i(T,\alpha)=E_{i0}(T)+E_{i2}(T)\,\alpha^2+E_{i4}(T)\,\alpha^4
\end{equation}\label{fitting_2}

\noindent with odd terms in $\alpha$ not included to respect the isospin
symmetry of the strong force. The symmetry energy is then given by

\begin{eqnarray}
E_{sym}(T,\rho)&=&E_{02}(T)+ E_{12}(T)\rho+ \nonumber
\\ &+&E_{22}(T)\rho^2+E_{32}(T)\rho^3
\end{eqnarray}

\noindent with the coefficients $E_{ij}(T)$ obtained from the fit of the CMD
data. In our case we calculate the symmetry energy separately for the
homogeneous and non-homogeneous cases. \\

Next, the tools presented in this Section will be used to analyze the
pasta structures, determine whether the phase transitions obtained
in Ref.~\cite{dorso2014} for symmetric nuclear matter survive in 
asymmetric nuclear matter, and what is the behavior of the symmetry 
energy within the pasta.

\subsection{Molecular dynamics simulation of nuclear matter}

In what follows, to study infinite systems of isospin-asymmetric nuclear matter,
the LAMMPS CMD code was fitted with the Pandharipande potentials and used to
track the evolution of systems with $A=$ 4000  or 6000 nucleons situated in a
cubic cell under periodic boundary conditions. The isospin content varied
from $x = 0.3$, $0.4$, and $0.5$, and densities between $0.05\,$fm$^{-3}$ to
$0.085\,$fm$^{-3}$. The temperature was controlled with a Nos\'e-Hoover
thermostat slowly varying from $T$ = 4 MeV down to 0.2 MeV ($\Delta T < 
0.1\%$). After placing the nucleons at random, but with a minimum 
inter-particle 
distance of 0.01 fm, the nucleons were endowed with velocities according to a 
Maxwell-Boltzmann distribution to correspond to a desired temperature, and the 
equations of motion were solved to mimic the evolution of the system. The 
nucleon positions, momenta, energy per nucleon, pressure, temperature, and 
density, were stored at fixed time-steps.\\

\section{Results for nuclear pasta}\label{results_1}

\subsection{\label{thermo}Thermodynamic magnitudes}

\begin{figure}
\begin{center}
   \includegraphics[width=3.5in]{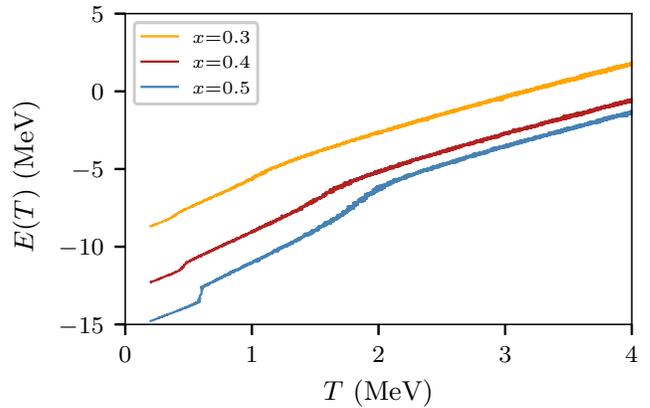}
\caption{The caloric curve for nuclear matter at
$\rho=0.05\,\mathrm{fm}^{-3}$ and $x = 0.3$, $0.4$, $0.5$, respectively. The
nucleons density was $\rho=0.05\,$fm$^{-3}$. The total number of nucleons was
6000.}\label{cc}
\end{center}
\end{figure}

\begin{figure}
\begin{center}
   \includegraphics[width=3.2in]{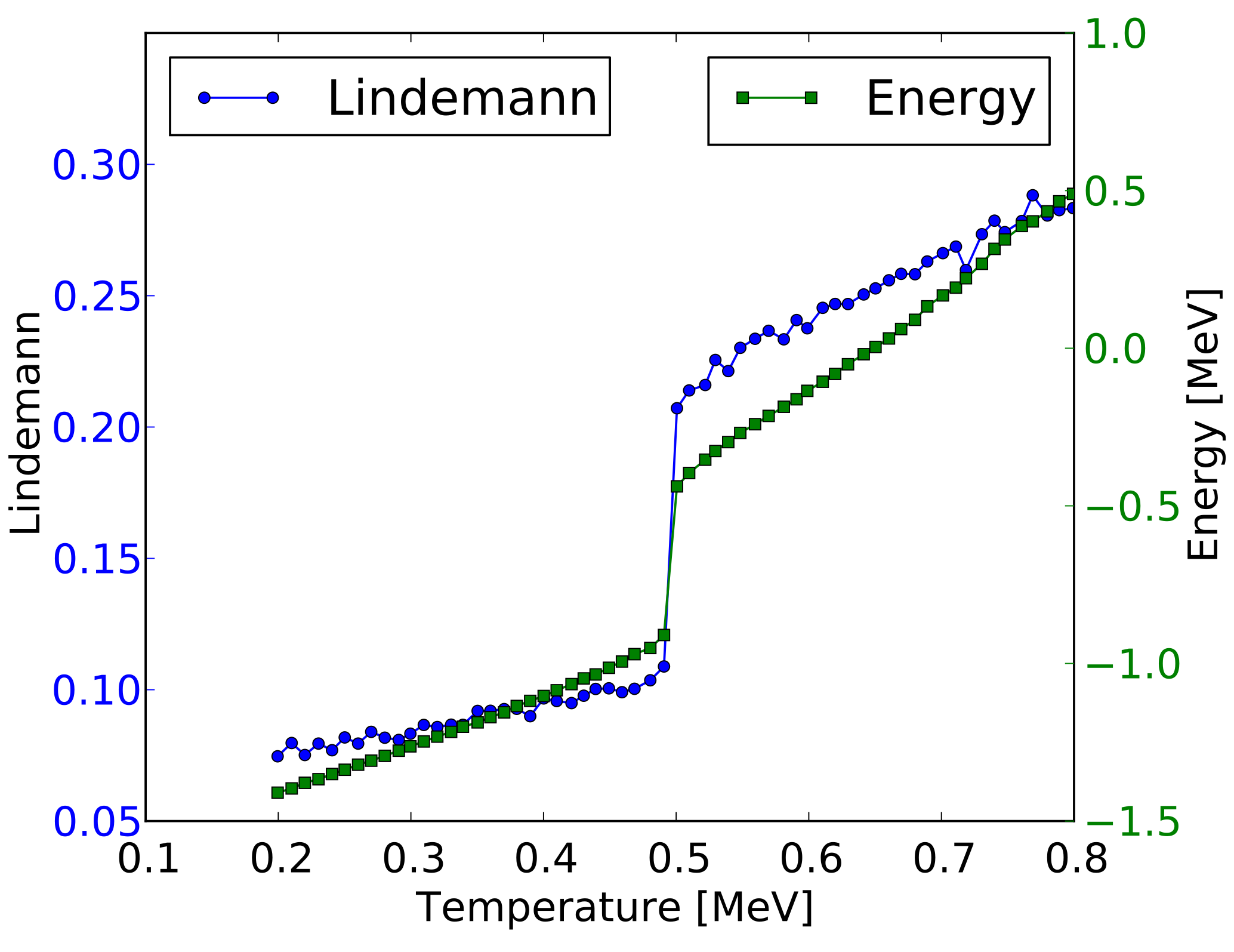}
\end{center}
\caption{Lindemann coefficient superimposed on the caloric
curve for $\rho = 0.05\,$fm$^{-3}$ for a symmetric system. The sudden change 
in going from $T<$ 0.5 MeV to $T>$ 0.5 MeV signals a phase change from solid 
pasta to liquid pasta. Extracted from Ref.~\cite{dorso2014}}\label{lin}.
\end{figure}

As a first step, we computed the caloric curve for three different proton
ratios. Figure~\ref{cc} shows the response at $\rho=0.05\,$fm$^{-3}$ 
for systems of 6000 nucleons at proton ratios of $x =$0.3, 0.4 and 0.5. The 
sharp 
change close to $T=0.5\,$MeV in symmetric nuclear matter was already examined 
in Ref.~\cite{dorso2014}. Figure~\ref{lin} shows an example of the behavior of 
the Lindemann coefficient as a function of the temperature, compared to the 
caloric 
curve for $\rho = 0.05\,$fm$^{-3}$. This change is a signal of a solid-liquid 
phase 
transition within the pasta regime, as concluded in Ref.~\cite{dorso2014}.  \\

Figure~\ref{cc} exhibits, however, a wider picture of the caloric responses.
The slope of the curves change gently at ``warm'' temperatures (say, near
$2\,$MeV) attaining lower energies than expected from the extrapolated values
for $T>2\,$MeV. It seems, though, that this change in the slope is more
significant as $\alpha\rightarrow 0$ ($x\rightarrow 0.5$).   \\

\begin{figure*}[!htbp]
\centering
\subfloat[$x=0.5$]{
\includegraphics[width=\columnwidth]
{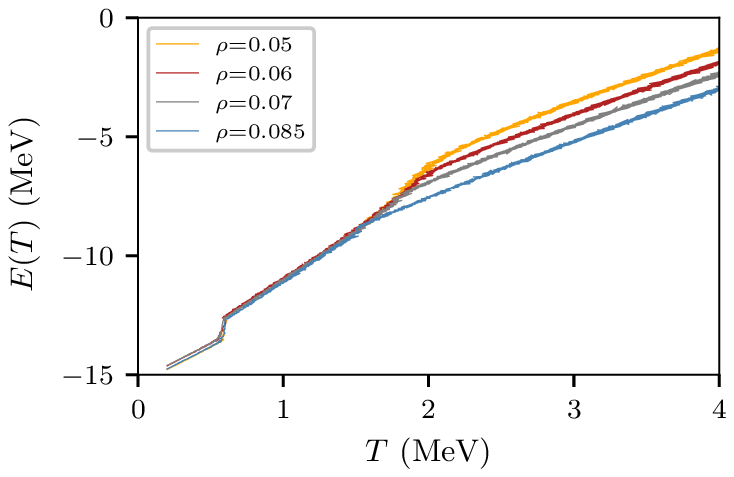}
} 
\subfloat[$x=0.5$]{
\includegraphics[width=\columnwidth]
{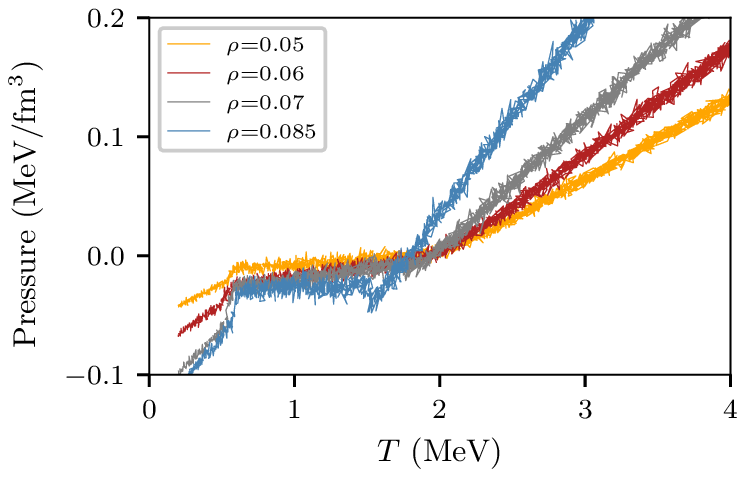}
} 
\\
\subfloat[$x=0.4$]{
\includegraphics[width=\columnwidth]
{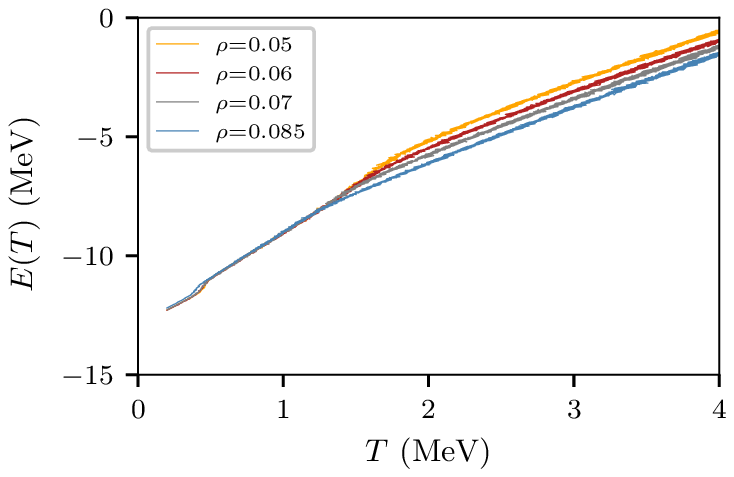}
} 
\subfloat[$x=0.4$]{
\includegraphics[width=\columnwidth]
{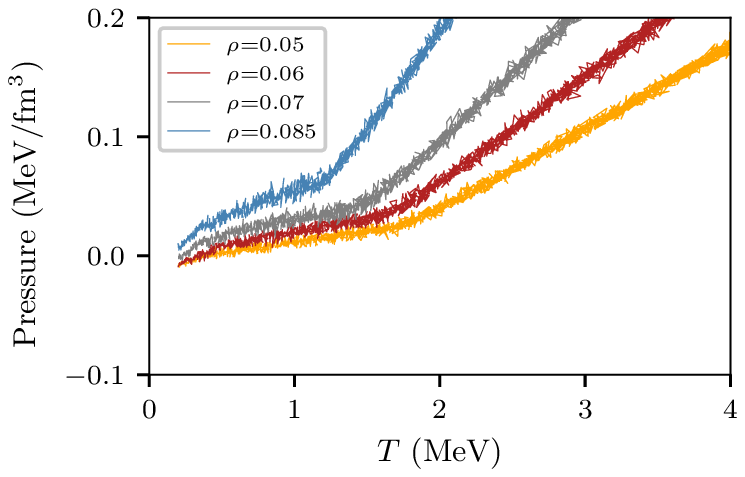}
}
\\
\subfloat[$x=0.3$]{
\includegraphics[width=\columnwidth]
{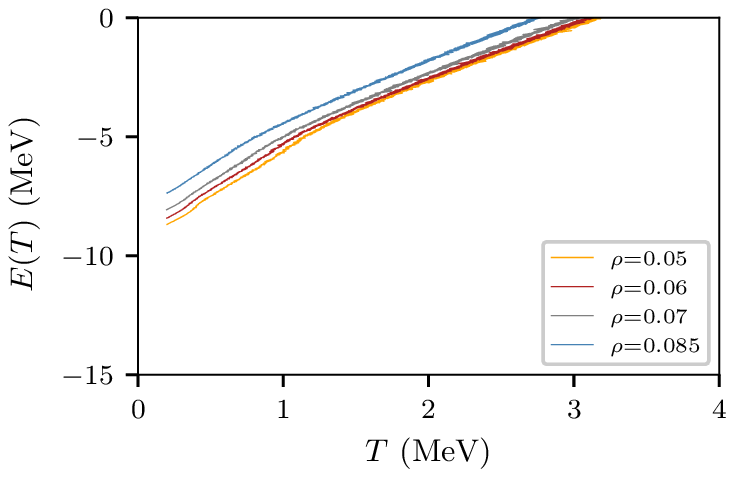}
} 
\subfloat[$x=0.3$]{
\includegraphics[width=\columnwidth]
{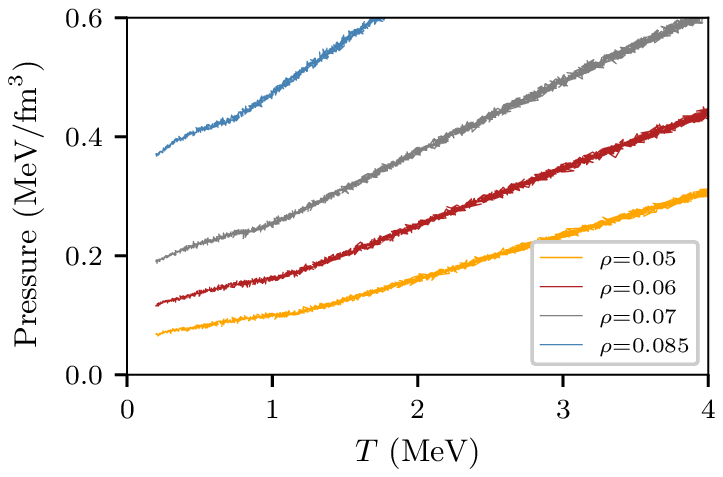}
}
\caption{\label{fig:energy_pressure} (Color online) Energy and
pressure as a function of temperature  for nuclear matter systems with 
6000 nucleons at densities $\rho$=0.05, 0.06, 0.07 and 
0.085$\,$fm${}^{-3}$.  Cases (a) and (b) are for symmetric nuclear matter 
($x=0.5$) and 
correspond to the configurations shown in Fig.~\ref{fig:pasta}, (c) and (d) are 
for 
non-symmetric nuclear matter with $x=0.4$, and (e) and (f) for $x=0.3$.}
\end{figure*}

The system pressure also experiences a change in the slope at similar
temperatures as the energy, as can be seen in Fig.~\ref{fig:energy_pressure}.
Notice that the pressure of the symmetric nuclear matter ($x=0.5$) also changes
sign, while the pressure of the asymmetric nuclear matter ($x<0.5$) remains
positive until very low temperatures. It is clear that the former enters into
the metastable regime, while the latter does not completely fulfill this
condition. \\

The change in the pressure sign is meaningful. Positive pressures may
be associated to  net repulsive inter-particle forces (regardless of momentum), 
while negative pressures may be associated to net attractive forces. The
visual image of a net attractive metastable scenario corresponds to the
\textit{pasta formation}.  The fact that pressure does not change sign for 
asymmetric nuclear matter
suggests a more complex scenario; Section \ref{microscopic} examines the low
density regime in detail.\\

\subsection{\label{microscopic}Microscopic magnitudes}

\begin{figure*}[!htbp]
\centering
\subfloat[$x=0.5$\label{rad_a}]{
\includegraphics[width=\columnwidth]
{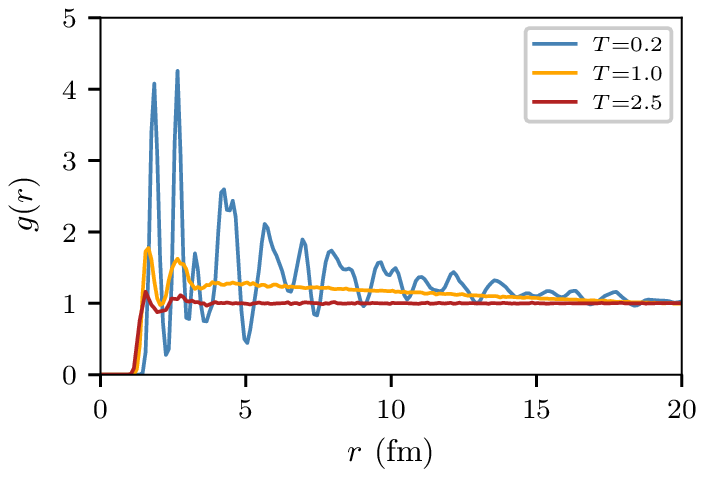}
} 
\subfloat[$x=0.4$\label{rad_b}]{
\includegraphics[width=\columnwidth]
{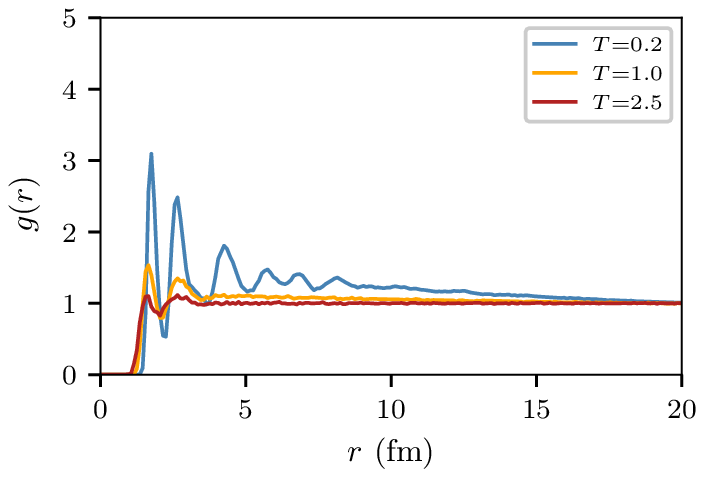}
} 
\caption{\label{rad} (Color online) Radial distribution
function $g(r)$ for the case of 6000 nucleons of symmetric (a) and 
non-symmetric 
(b) nuclear matter, at $\rho= 0.085\,$fm$^{-3}$ and $T = 0.2\,$MeV, $1.0\,$MeV 
and 2.5$\,$MeV. The binning is $0.1\,$fm in width in a simulation cell of 
41.2$\,$fm 
of width.}
\end{figure*}

We next explored the radial distribution function $g(r)$ (see Section 
\ref{pair}). Figure~\ref{rad} shows the $g(r)$ for nuclear matter at 
$\rho=0.085\,$fm$^{-3}$ and at three different temperatures (see caption for 
details). Panel (a) show the results for the symmetric ($x=0.5$) case, and (b) 
the non-symmetric ($x=0.4$) case.\\

From a very first inspection of Fig.~\ref{rad} it is clear that the
$T=0.2\,$MeV situation corresponds to a solid-like pattern. The local maxima
 indicate the (mean) equilibrium distance between nucleons in a orderly scheme.
 The first two maxima near the origin appear sharply, meaning well-arranged
neighbors around each nucleon. These maxima are still present at $T=1\,$MeV and
$T=2.5\,$MeV, although not as sharp as in the solid-like situation. Thus, the
very short range $r<5\,$fm seems not to present an unusual behavior,
 regardless of the first order transition (solid-to-liquid like) detailed in
Section \ref{thermo}, indicating that in spite of the phase change the pasta 
structure is maintained.\\

A noticeable different pattern can be seen in Fig.~\ref{rad_a} between the
$T=1\,$MeV  (metastable) and the $T=2.5\,$MeV (stable) scenario for the range
$r>5\,$fm. The metastable scenario exhibits a negative smooth slope, while the
stable one displays a vanishing slope. This is in agreement with the
pasta formation, as follows. \\

The $g(r)$ computation involves tallying distances between neighboring
nucleons \textit{at} the nucleons positions (see Eq.~(\ref{eqn:gr_def})).
Hence, the $g(r)$ widely samples the inner environment of the pasta,
at least for short distances. As the sampled distance increases (inside the
pasta), the $g(r)$ is not expected to converge to unity but above
unity, since the local average density exceeds the global average
density $N/V$. That is what we see in Fig.~\ref{rad_a} close to $5\,$fm.  \\

The above reasoning remains valid until the neighboring distances surpasses the
inner environment of the pasta. As bubbles or tunnels, or other hollow
spaces appear, the tally in Eq.~(\ref{eqn:gr_def}) converges to the total
number of nucleons inside the simulating cell. Consequently, the distribution
$g(r)$ diminishes. A precise computation of $g(r)$ for non-homogeneous 
media can be seen in Appendix~\ref{sec:gr_example}, together with some 
theoretical support.
\\

Notice from Fig.~\ref{rad}a that the solid-like situation also resembles the
pasta formation since the moving average of the $g(r)$ for
$T=0.2\,$MeV exhibits a smooth negative slope. \\

\begin{figure}  
\begin{center}
   \includegraphics[width=3.5in]{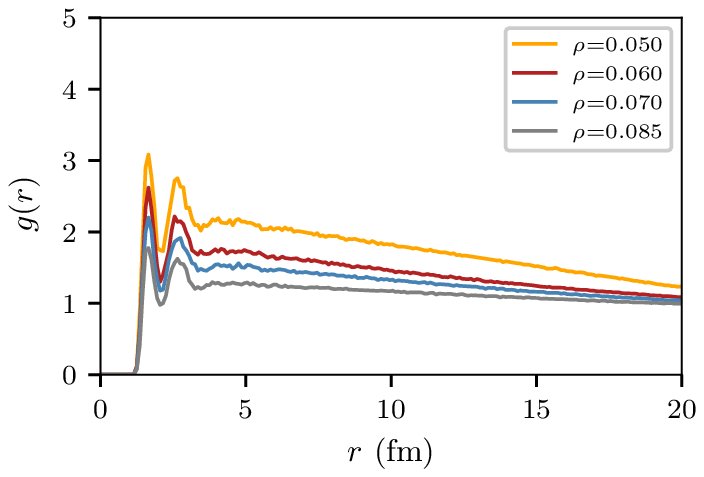}
\caption{(Color online) Radial distribution function $g(r)$ for the case
of 6000 nucleons at $T=1.0\,$MeV and
$x=0.5$. The densities are $\rho= 0.05$ (top curve), $0.06$, $0.07$ and 
$0.085\,$fm$^{-3}$ (bottom curve).
The binning is $0.1\,$fm width, while the simulation cell is approximately 
40$\,$fm width.
}\label{rad2}
\end{center}
\end{figure}

The argument outlined so far does not hypothesize on the pasta shape.
It only compares the average density inside the pasta with respect to
$N/V$. Thus, it is expected to hold on a variety of \textit{pastas}. This was
verified for different densities, as shown in Fig.~\ref{rad2}. \\

The non-symmetric case shown in Fig.~\ref{rad}b is somewhat different. The
long range $r>5\,$fm appears flattened with respect to the symmetric case
(Fig.~\ref{rad}a). Thus, no clear indication of pasta formation
can be seen here. The same happened, indeed,  with respect to the metastable
state in Section \ref{thermo}. Both, the pressure and the radial distribution
seem to fail in order to detect pasta structures.  \\

\begin{figure*}[!htbp]
\centering
\subfloat[\label{fig:rho085_all}]{
\includegraphics[width=0.65\columnwidth]
{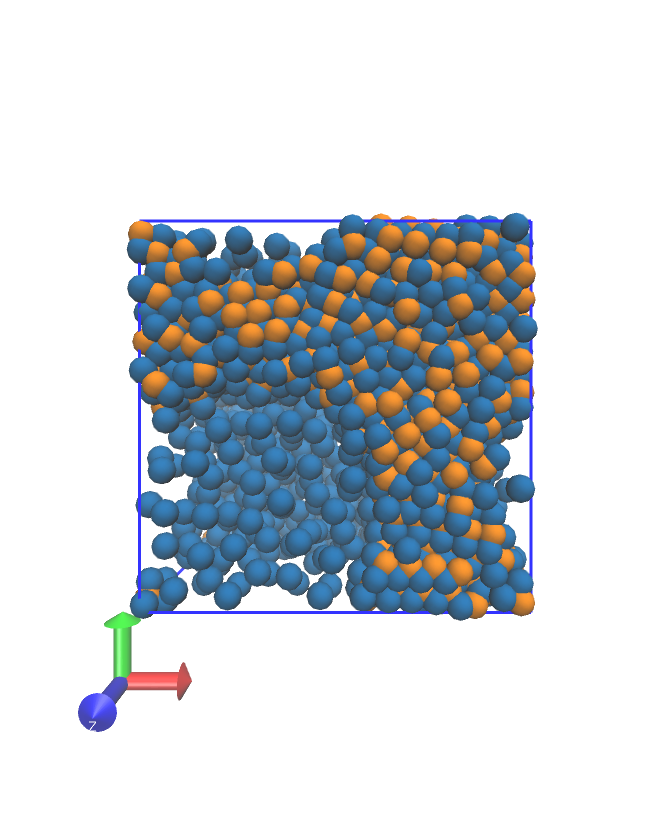}
}
\subfloat[\label{fig:rho085_protons}]{
\includegraphics[width=0.65\columnwidth]
{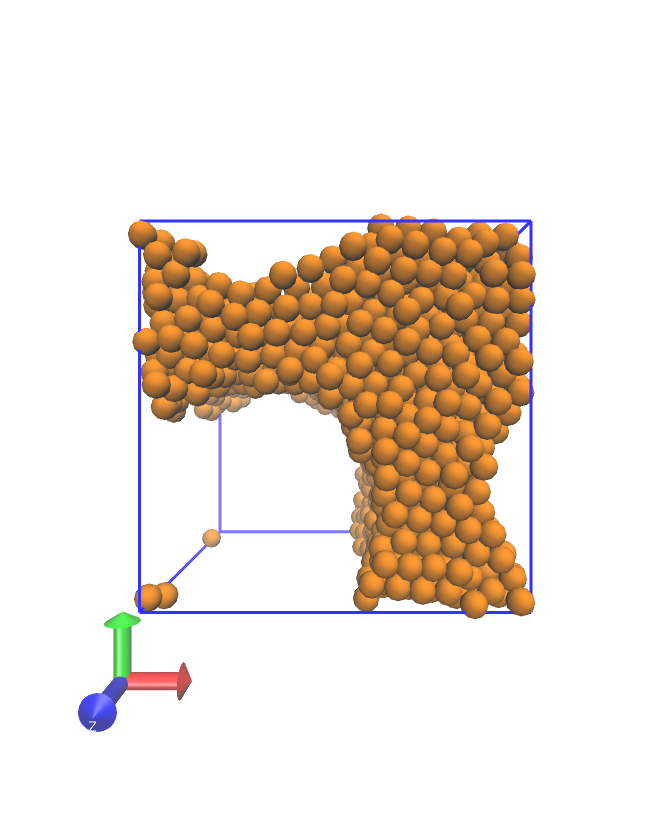}
}
\subfloat[\label{fig:rho085_neutrons}]{
\includegraphics[width=0.65\columnwidth]
{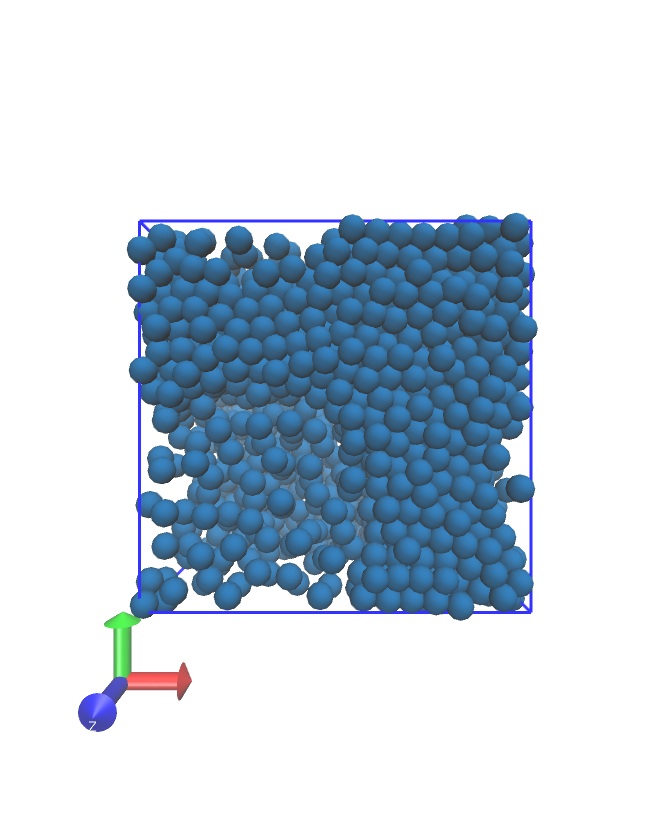}
}
\caption{\label{fig:pasta_decomposed} (Color online) Pasta structures for
nuclear matter systems with 6000 nucleons,
$x=0.4$ at $T=0.2\,$MeV and density $\rho$=0.085$\,$fm${}^{-3}$. (a) All
nucleons (protons in light color and neutrons in dark), (b) protons only, and 
(c)
neutrons only.}
\end{figure*}

\begin{figure*}[!htbp]
\centering
\subfloat[\label{fig:rho085_all_2}]{
\includegraphics[width=0.65\columnwidth]
{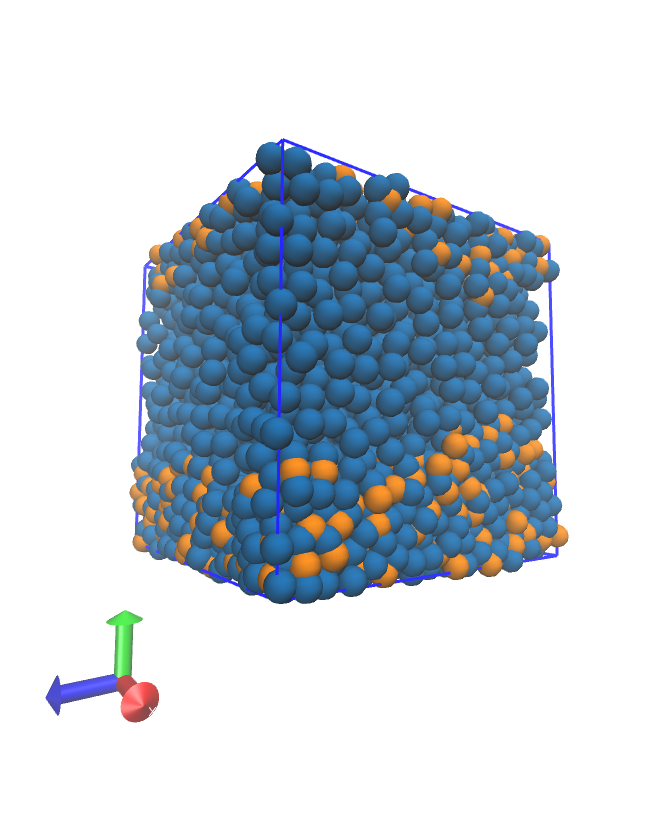}
}
\subfloat[\label{fig:rho085_protons_2}]{
\includegraphics[width=0.65\columnwidth]
{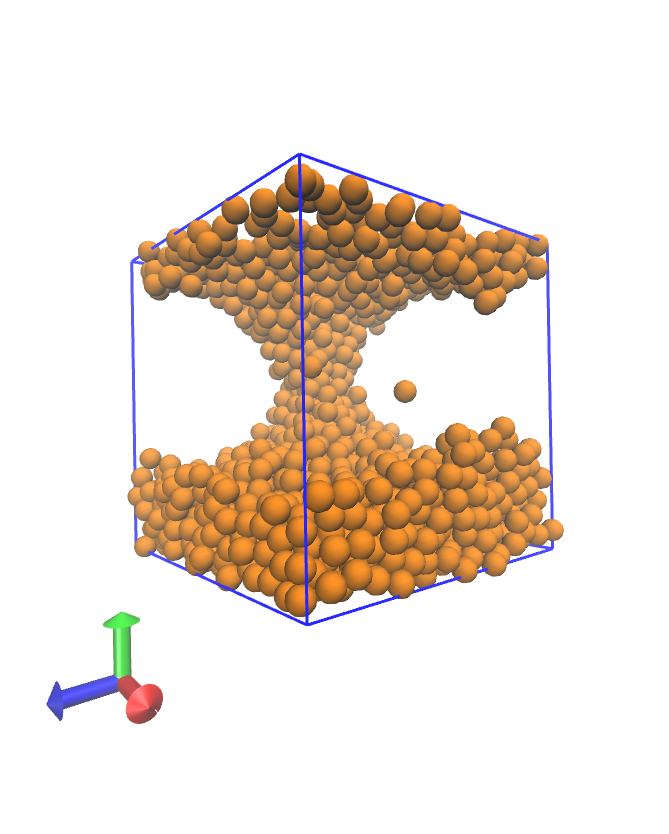}
}
\subfloat[\label{fig:rho085_neutrons_2}]{
\includegraphics[width=0.65\columnwidth]
{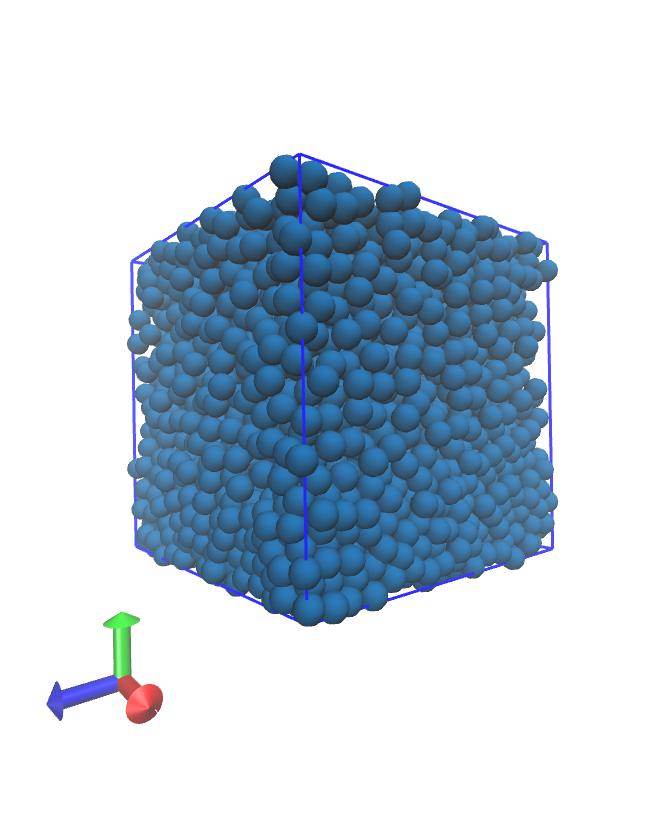}
}
\caption{\label{fig:pasta_decomposed_2} (Color online) Pasta structures
for nuclear matter systems with 6000 nucleons,
$x=0.3$ at $T=0.2\,$MeV and density $\rho$=0.085$\,$fm${}^{-3}$. (a) All
nucleons (protons in light color and neutrons in dark), (b) protons only, and 
(c)
neutrons only.}
\end{figure*}

The visual inspection of the simulation cell shows that pasta
structures are actually present in non-symmetric nuclear matter.
Figs.~\ref{fig:pasta_decomposed} and \ref{fig:pasta_decomposed_2} exhibit two
non-symmetric situations. Figure~\ref{fig:pasta_decomposed} actually
corresponds to the same configurations as in Fig.~\ref{rad}b but now protons and
neutrons are represented separately. The protons actually exhibit a sharp
pasta formation (see Fig.~~\ref{fig:rho085_protons}). The neutrons'
pattern, instead, appears fuzzy because of the fraction in excess that occupies
the empty volume left by the pasta.  \\

We realize from the above observations that many indicators may fail to
recognize the pasta formation in non-symmetric nuclear matter. This is
because the neutrons in excess obscure the useful information when mean values
are computed. This drawback may also occur in symmetric nuclear matter at the
very beginning of the pasta formation (say, at $T\sim 2\,$MeV) since
the presence of small bubbles or tiny tunnels may be lost in the averaging
process. Section \ref{morphology} deals with these issues. \\

\subsection{\label{morphology}Morphological indicators}

In order to capture the early pasta formation at ``warm'' temperatures (say, 
$T\sim 2\,$MeV) we introduce two indicators that do not require computing mean 
values, namely the Kolmogorov statistic and the Minkowski functionals; the 
former is not conditioned to space binning, as the latter is, but they both 
provide different points of view on the pasta formation. \\

\subsubsection{\label{kolmogorov_results} Kolmogorov statistic}

The Kolmogorov statistic is known to be ``distribution-free'' for the
univariate distributions case and ``almost distribution-free'' for multivariate
distributions (see Section~\ref{kolm}). This section deals with univariate
distributions, while the multivariate distributions are left to Section
\ref{subsec:esym}.\\

\begin{figure*}[!htbp]
\centering
\subfloat[\label{fig:rho05_x05_t20}]{
\includegraphics[width=3.5in]
{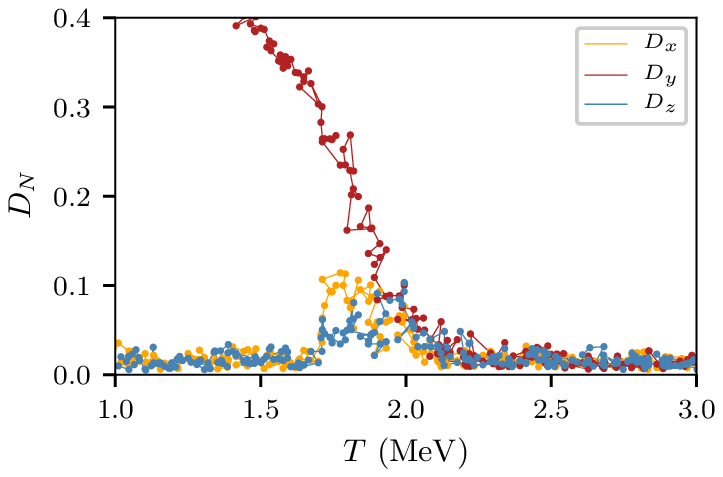}
}
\subfloat[\label{fig:rho05_x05_t20_image}]{
\includegraphics[width=0.7\columnwidth]
{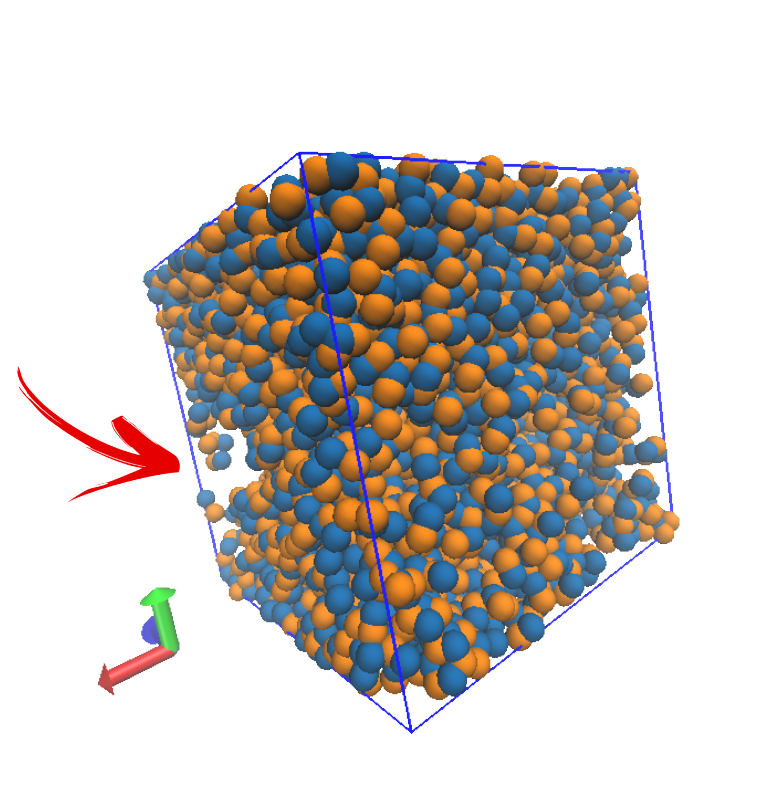}
}
\caption{\label{fig:rho05_x05_t20_all} (Color online) (a) The Kolmogorov
1D statistic vs. temperature. Data corresponds to the position of 6000 nucleons
 at $\rho=0.05\,\mathrm{fm}^{-3}$ and $x = 0.5$ (the same configuration as in
Fig.~\ref{fig:rho05}). $D_x$, $D_y$ and $D_z$ correspond to the $x$, $y$ and $z$
Kolmogorov statistics, respectively (see text for details). (b) Visualization of
the system analyzed in (a) at $T=2\,$MeV. The arrow points to the most
noticeable bubble appearing in the picture. }
\end{figure*}

The 1D Kolmogorov statistic is useful to study the departure of the
nucleons' position from homogeneity, that is, from the uniform distribution.
We applied the statistic separately on each cartesian coordinate, this may be 
envisaged as the projection of each position on the three
canonical axes. Figure~\ref{fig:rho05_x05_t20_all}  shows the corresponding
results for an isospin symmetric system at $\rho = 0.05\,$fm$^{-3}$. \\

As can be seen in Fig.~\ref{fig:rho05_x05_t20_all}, the values of the
1D Kolmogorov statistic (\textit{i.e.} the discrepancy) are negligible for
temperatures above $2\,$MeV, as expected for energetic particles moving around
homogeneously. At $T\approx 2\,$MeV, the three statistics experience a change in
the slope, although $D_x$ and $D_z$ return to negligible values as the
temperature further decreases. The $D_y$ statistic, instead, attains a definite
departure from homogeneity for $T<2\,$MeV. Recall from Fig.~\ref{fig:rho05} that
a \textit{lasagna} or slab-like structure occurs across the $y$-axis as
temperature is lowered.  \\

\begin{figure*}[!htbp]
\centering
\subfloat[\label{fig:rho05_x05_T2_0}$T=2\,$MeV]{
\includegraphics[width=0.6\columnwidth]
{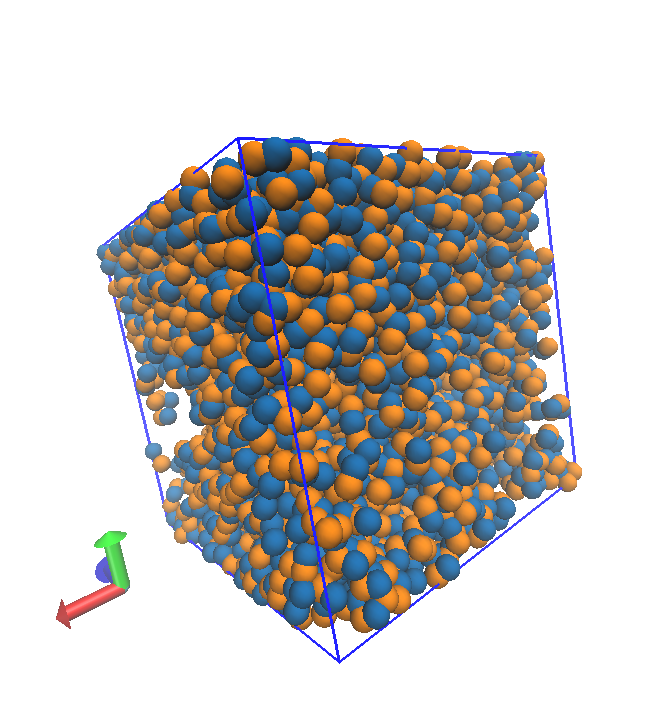}
}
\subfloat[\label{fig:rho05_x05_T1_8}$T=1.8\,$MeV]{
\includegraphics[width=0.6\columnwidth]
{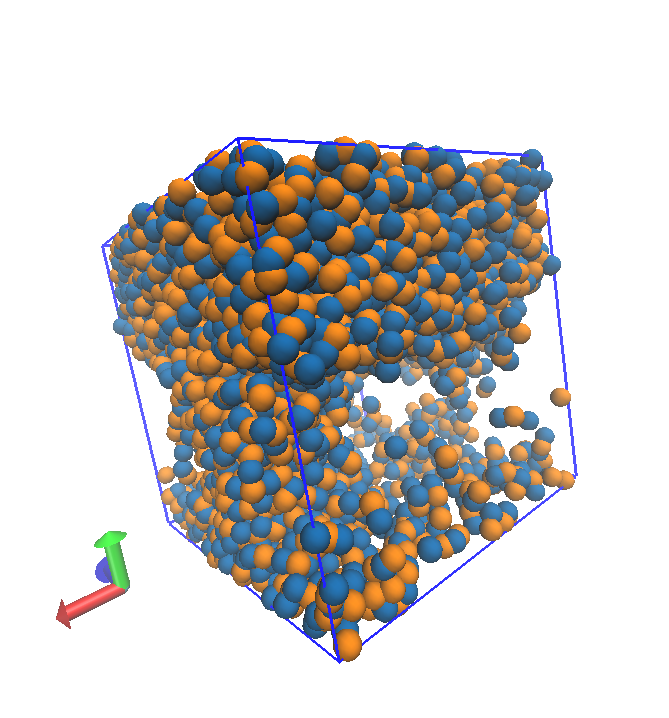}
}
\subfloat[\label{fig:rho05_x05_T1_5}$T=1.5\,$MeV]{
\includegraphics[width=0.6\columnwidth]
{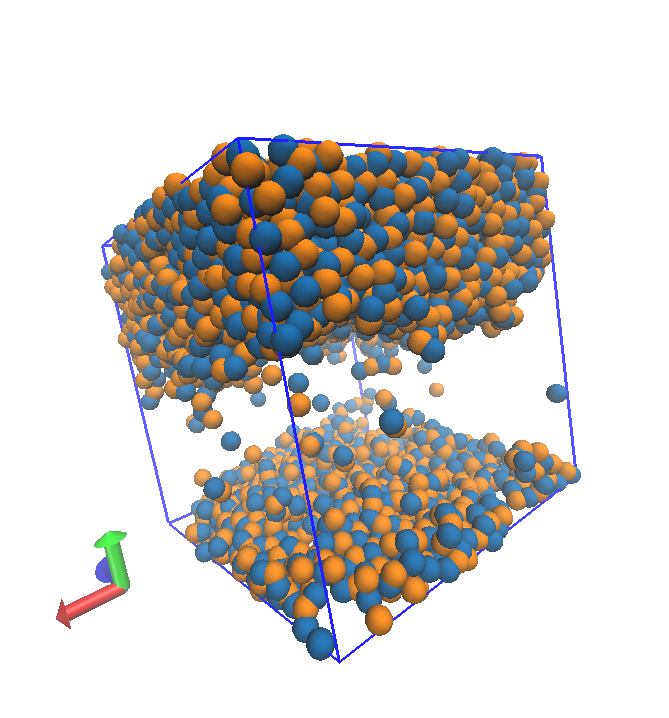}
}
\caption{\label{fig:rho05_x05_seq} (Color online) Visualization of 6000
nucleons  at $\rho=0.05\,\mathrm{fm}^{-3}$, $x = 0.5$ (the same configuration
as in Fig.~\ref{fig:rho05}) at three temperatures. (a) A bubble can be seen on 
the left. (b) A
tunnel appears along the $x,y$ plane. (c) The tunnel widens and breaks into two
slabs.  }
\end{figure*}

The 1D Kolmogorov statistic attains the departure from homogeneity at an early
stage of the pasta formation. The arrow in
Fig.~\ref{fig:rho05_x05_t20_image} points to the most noticeable bubble
appearing in the system at $T=2\,$MeV. The bubble-like heterogeneity
also explains the changes in the slope for $D_x$ and $D_z$ at this temperature,
as pictured in Fig.~\ref{fig:rho05_x05_seq}. For decreasing temperatures, the
bubble widens (on the right side of the image due to the periodic boundary
conditions), a tunnel appears (Fig.~\ref{fig:rho05_x05_T1_8}), and at 
$T<1.5\,$MeV 
it finally splits into two pieces while the $(x,y)$ homogeneity gets
restored (Fig.~\ref{fig:rho05_x05_T1_5}) returning $D_x$ and 
$D_z$ back to their negligible values.\\

\begin{figure*}[!htbp]
\centering
\subfloat[\label{fig:rho05_pn}$\rho=0.05\,\mathrm{fm}^{-3}$]{
\includegraphics[width=\columnwidth]
{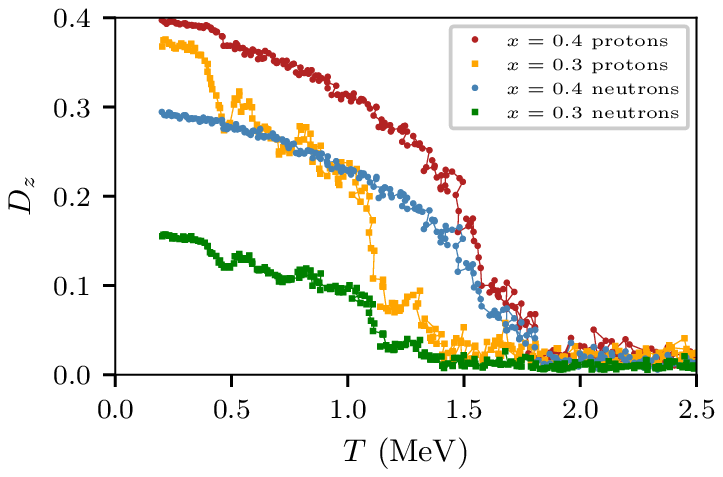}
}
\subfloat[\label{fig:rho085_pn}$\rho=0.085\,\mathrm{fm}^{-3}$]{
\includegraphics[width=\columnwidth]
{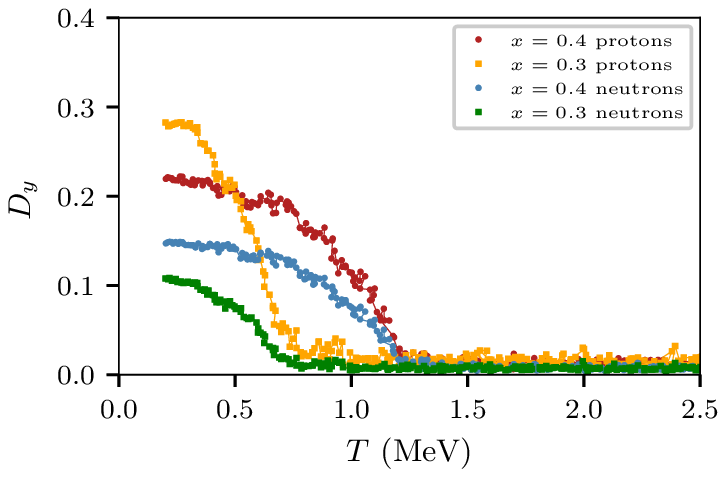}
}
\caption{\label{fig:Kolmogorov_asym} (Color online) (a) The Kolmogorov
1D statistic vs. temperature. Data corresponds to the position of 6000 nucleons
(no screened Coulomb potential) for $x = 0.3$ and $0.4$. (a) The $D_z$
statistic ($\rho=0.05\,\mathrm{fm}^{-3}$). (b) The $D_y$ statistic
($\rho=0.085\,\mathrm{fm}^{-3}$). }
\end{figure*}

\begin{figure*}[!htbp]
\centering
\subfloat[\label{fig:rho05_all}]{
\includegraphics[width=0.65\columnwidth]
{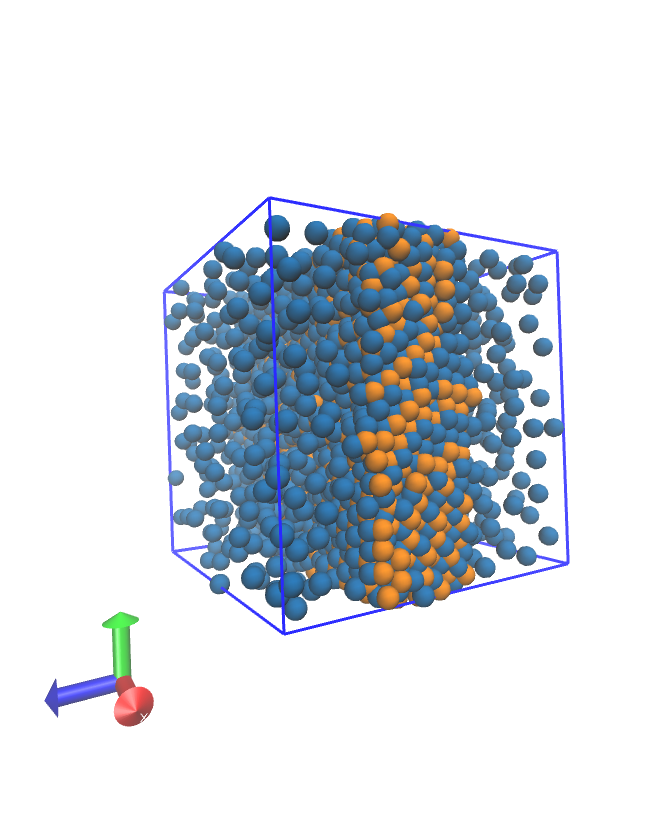}
}
\subfloat[\label{fig:rho05_protons}]{
\includegraphics[width=0.65\columnwidth]
{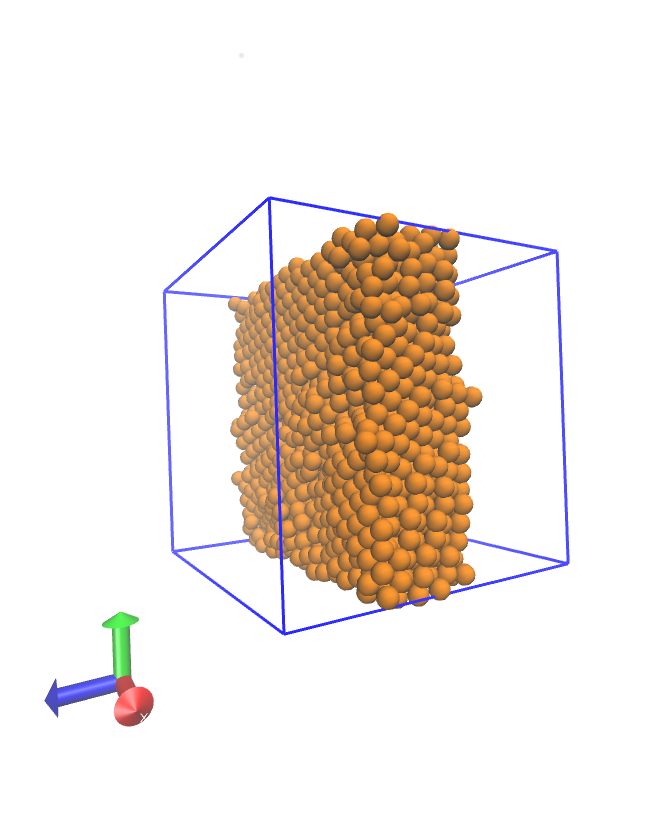}
}
\subfloat[\label{fig:rho05_neutrons}]{
\includegraphics[width=0.65\columnwidth]
{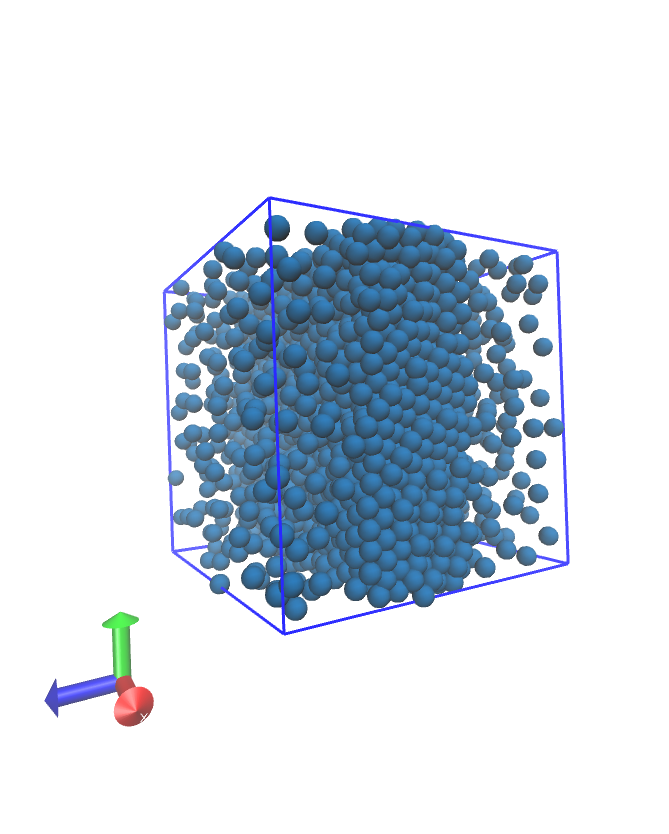}
}
\caption{\label{fig:pasta_decomposed_3} (Color online) Pasta structures
for nuclear matter systems with 6000 nucleons,
$x=0.4$ at $T=0.2\,$MeV and density $\rho$=0.05$\,$fm${}^{-3}$. (a) All
nucleons (protons in orange and neutrons in blue). (b) Protons only. (c)
Neutrons only.}
\end{figure*}

\begin{figure*}[!htbp]
\centering
\subfloat[\label{fig:rho05_all_2}]{
\includegraphics[width=0.65\columnwidth]
{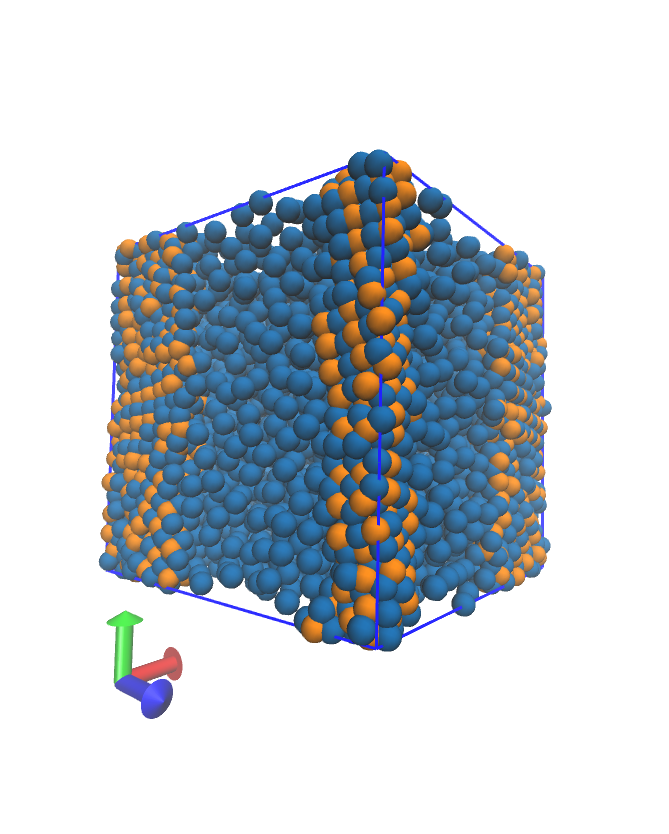}
}
\subfloat[\label{fig:rho05_protons_2}]{
\includegraphics[width=0.75\columnwidth]
{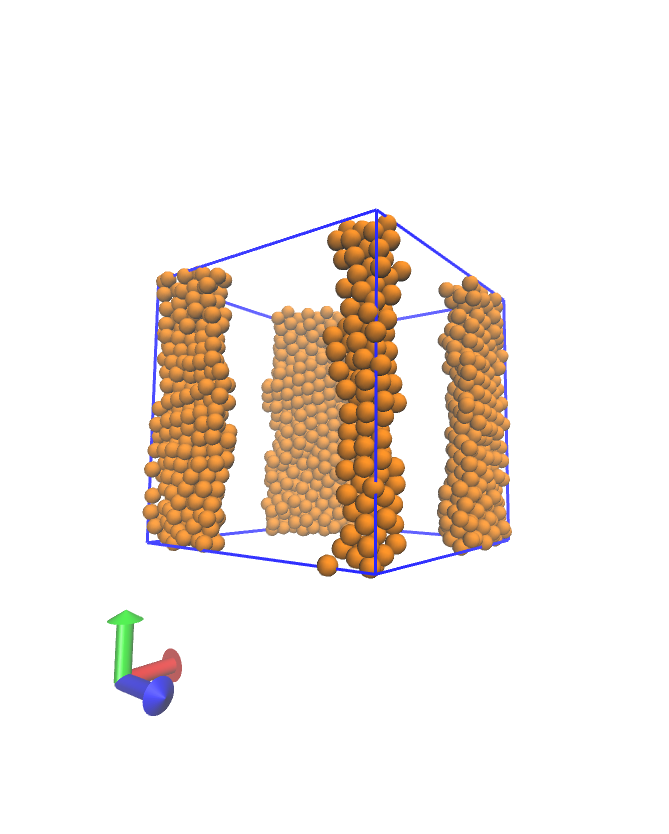}
}
\subfloat[\label{fig:rho05_neutrons_2}]{
\includegraphics[width=0.7\columnwidth]
{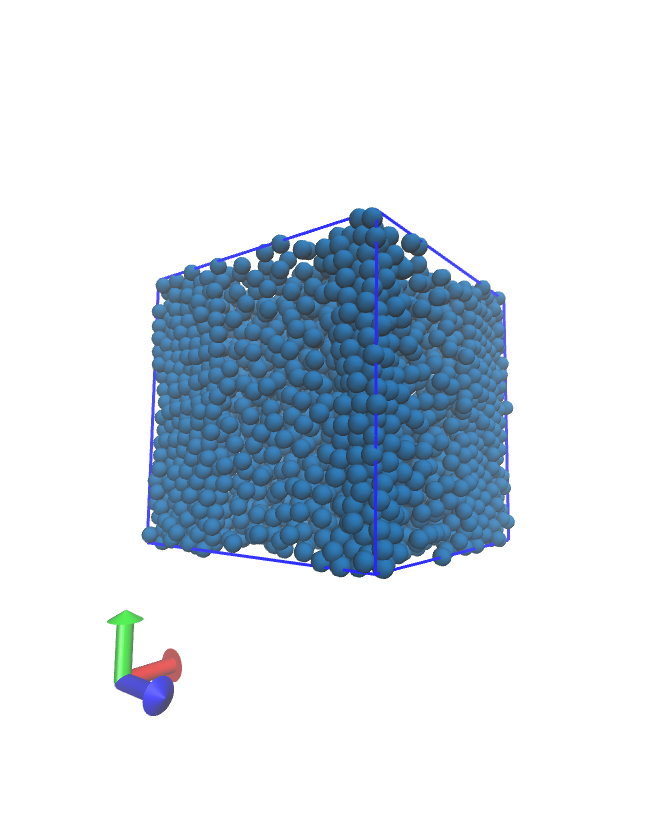}
}
\caption{\label{fig:pasta_decomposed_4} (Color online) Pasta structures
for nuclear matter systems with 6000 nucleons,
$x=0.3$ at $T=0.2\,$MeV and density $\rho$=0.05$\,$fm${}^{-3}$. (a) All
nucleons (protons in orange and neutrons in blue). (b) Protons only. (c)
Neutrons only.}
\end{figure*}

We further applied the 1D Kolmogorov statistic on protons and neutrons
separately for non-symmetric nuclear matter systems. Figure
\ref{fig:Kolmogorov_asym} exhibits the most significant 1D statistics for the
densities $\rho=0.05\,\mathrm{fm}^{-3}$ and $\rho=0.085\,\mathrm{fm}^{-3}$,
respectively. The corresponding spatial configurations can be seen in
Figs.~\ref{fig:pasta_decomposed} and \ref{fig:pasta_decomposed_2} for 
$\rho=0.085\,\mathrm{fm}^{-3}$,
and in Figs.~\ref{fig:pasta_decomposed_3} and \ref{fig:pasta_decomposed_4} for 
$\rho=0.05\,\mathrm{fm}^{-3}$.\\

According to Fig.~\ref{fig:Kolmogorov_asym}, the temperature threshold at which
the 1D Kolmogorov statistic becomes significant, decreases with smaller proton
ratios (and fixed density). This is in agreement with the caloric curves
presented in Section \ref{thermo}, but more sharply exposed now. However, both
species, protons and neutrons, seem to depart from homogeneity at the same
temperature threshold (see Fig.~\ref{fig:Kolmogorov_asym}). The departure
appears more sharply for the protons than for the neutrons. Indeed, the protons
attain higher maximum values of $D$ than the neutrons at low temperatures (for
the same configuration). The visual inspection of
Figs.~\ref{fig:pasta_decomposed}, \ref{fig:pasta_decomposed_2}, 
\ref{fig:pasta_decomposed_3} and \ref{fig:pasta_decomposed_4} confirms this
point.  \\

We conclude that the non-symmetric systems do not
develop noticeable bubbles or heterogeneities at the same temperature 
as the symmetric systems do. The excess of neutrons 
appears to frustrate the pasta formation for a while, but the
protons manage to form pasta structures at lower temperatures. The released 
neutrons get distributed along the cell disrupting the pasta structure, as 
mentioned in Section \ref{microscopic}. \\

\subsubsection{\label{minkowski_results} Minkowski functionals}

The Minkowski functionals supply complementary information on the
pasta structure from the early stage to the solid-like stage. The
accuracy of this information is, however, conditional to the correct binning of
the simulation cell. Tiny ``voxels'' (that is, a high density binning)
may produce fake empty voids (artificial bubbles or tunnels) and, on the
contrary, oversized voxels may yield a wrong structure of the system 
due to the lack of details. Therefore, some effort needs to be spent to 
determine the correct size for the voxels; Appendix \ref{sec:voxel} 
summarizes this procedure.  \\

\begin{figure}  
\begin{center}
   \includegraphics[width=\columnwidth]{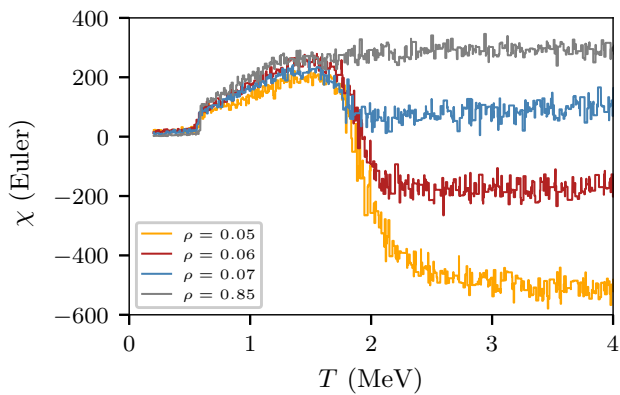}
\caption{(Color online) The Euler functional $\chi$ as a function of
temperature for the densities $\rho=0.05$, $0.06$, $0.07$, $0.085\,$fm$^{-3}$
and the proton ratio $x=0.5$. The voxel edge length is $d=2.35\,$fm. The
total number of nucleons is $6000$.}\label{mink_x05}
\end{center}
\end{figure}

The simulation cell was first divided into cubic voxels of edge length
$d=2.35\,$fm. The Euler functional $\chi$ was computed for symmetric nuclear
matter, according to Eq.~(\ref{eq:chi}) and the results are shown in
Fig.~\ref{mink_x05}.  As seen in this figure, the Euler functional $\chi$ for 
symmetric matter 
shows three distinct  regions as a function of temperature, one at $T>2\,$MeV, 
one 
at $T<0.5\,$MeV, and a transition one between these two. \\

\begin{figure*}[!htbp]
\centering
\subfloat[$\rho=0.05$]{
\includegraphics[width=0.5\columnwidth]
{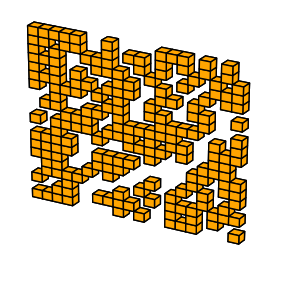}
} 
\subfloat[$\rho=0.06$]{
\includegraphics[width=0.5\columnwidth]
{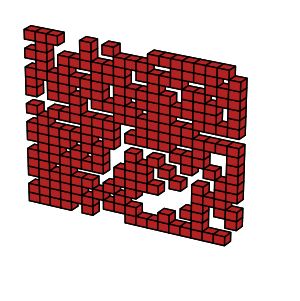}
} 
\subfloat[$\rho=0.07$]{
\includegraphics[width=0.5\columnwidth]
{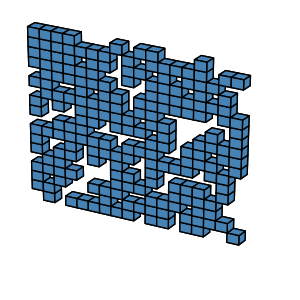}
} 
\subfloat[$\rho=0.085$]{
\includegraphics[width=0.5\columnwidth]
{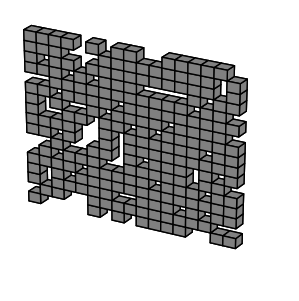}
} 
\caption{\label{fig:mink_slice} (Color online) Inside view of discretized
nuclear matter with 6000 nucleons at
$T=2.5\,$MeV and $x=0.5$. The discretization turned the 6000 nucleons into
(approximately) 4300 voxels (edge length $d=2.35\,$fm). Only a slice from the
middle ($y$-plane) of the simulation cell is represented. The labels correspond
to the density values in fm$^{-3}$. The colors are in correspondence with
Fig.~\ref{mink_x05}.}
\end{figure*}

At $T>2\,$MeV $\chi$ does not experience significant changes although it 
attains 
different values depending on the density. At the low densities of 
$\rho=0.05\,$fm$^{-3}$ and $\rho=0.06\,$fm$^{-3}$ $\chi$ exhibits negative 
values which, according to Eq.~(\ref{eq:chi}), indicate that the nucleons are 
sparse 
enough to form tunnels and empty regions across the cell. At higher densities, 
however, $\chi$ becomes positive indicating that tunnels begin to fill forming 
cavities 
and isolated regions. This is confirmed by Fig.~\ref{fig:mink_slice} which shows 
an 
inside view of the discretized nuclear matter at $T=2.5\,$MeV, $x=0.5$ and for 
the 
four densities under consideration.  \\

The complementarity of $\chi$ over other measures can be seen by comparing 
it to, for instance, the results of the Kolmogorov statistics 
(Section~\ref{kolmogorov_results}). 
As seen in Section~\ref{subsec:esym}, as the nucleons get distributed 
uniformly at high temperatures neither the 1D nor the 3D Kolmogorov statistic 
capture the qualitative difference between a tunnel-like and a cavity-like 
scenarios. 
Both landscapes may not exhibit noticeable heterogeneities, and thus, they 
appear to
be essentially the same from the point of view of the Kolmogorov statistic. \\

The energy, however, distinguishes between aforementioned different scenarios.
From the comparison between the caloric curves introduced in
Section~\ref{thermo} and the current Euler functional, one can see that both
magnitudes are density-dependent in the high temperature regime. While
$\chi$ increases for increasing densities, the (mean) energy per nucleon
diminishes (see Figs.~\ref{fig:energy_pressure} and~\ref{mink_x05}). Thus,
the less energetic configuration (say, $\rho=0.085\,$fm$^{-3}$) appears to be
a cavity-like (or small bubble-like) scenario from the point of view of $\chi$ 
(with voxels size of $d=2.35\,$fm). \\

The Euler functional $\chi$ exhibits a dramatic change at $T\approx 2\,$MeV. 
This is associated to the departure from homogeneity at the early stage of 
the pasta formation, as already mentioned in Section~\ref{minkowski_results}. 
Notice that the $\chi$ values for the examined densities join into a 
single pattern for $T<2\,$MeV, in agreement with the behavior of the energy 
seen in Fig.~\ref{fig:energy_pressure}.  \\

\begin{figure*}[!htbp]
\centering
\subfloat[$\rho=0.05$]{
\includegraphics[width=0.6\columnwidth]
{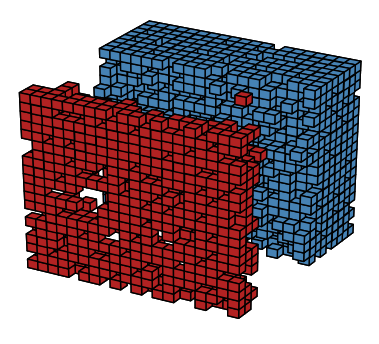}
} 
\hspace{3cm}\subfloat[$\rho=0.085$]{
\includegraphics[width=0.6\columnwidth]
{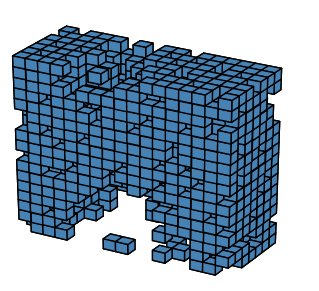}
} 
\caption{\label{fig:mink_slice_2} (Color online) Inside view of
discretized nuclear matter with 6000 nucleons
at $T=1.5\,$MeV and $x=0.5$. The discretization turned the 6000 nucleons into
(approximately) 4300 voxels (edge length $d=2.35\,$fm). (a) Density
$\rho=0.05\,$fm$^{-3}$. The system split into two major pieces, represented
in different colors. Many small pieces in the middle gap can not be seen.
(b) Density $\rho=0.085\,$fm$^{-3}$. Only a mid-slice of seven voxels thick
is shown for practical reasons. }
\end{figure*}

It should be emphasized that although all the examined densities share the same
$\chi$ pattern for $T<2\,$MeV, their current morphology may be quite
different. Fig.~\ref{fig:mink_slice_2} illustrates two such situations
(see caption for details). It seems, though, that whatever the morphology,
these are constrained to be equally energetic (see 
Fig.~\ref{fig:energy_pressure}). \\

\begin{figure*}[!htbp]
\centering
\subfloat[$\rho=0.05$\label{fig:mink_func_a}]{
\includegraphics[width=\columnwidth]
{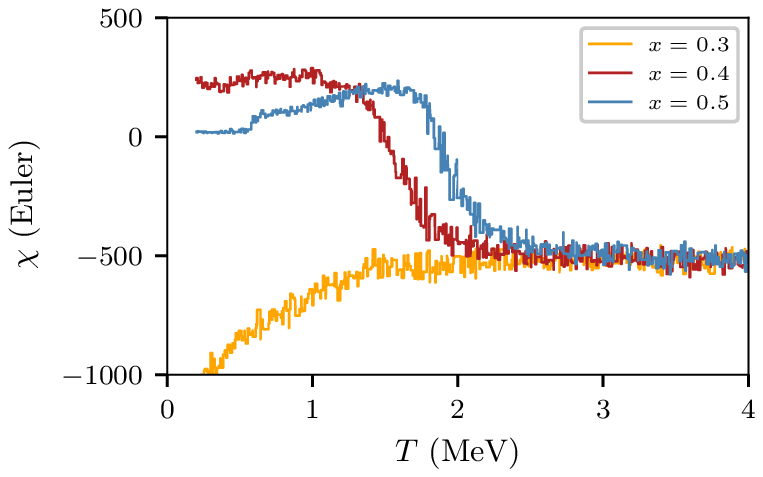}
} 
\subfloat[$\rho=0.085$\label{fig:mink_func_b}]{
\includegraphics[width=\columnwidth]
{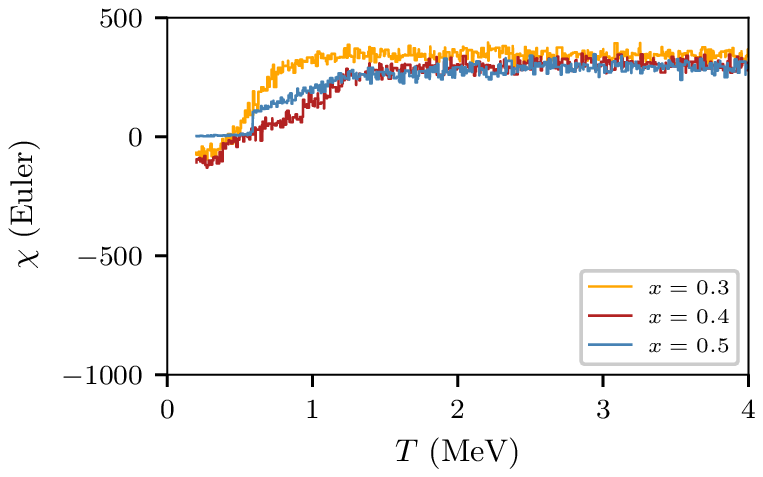}
} 
\caption{\label{fig:mink_func} (Color online) The Euler functional as
a function of temperature for $x=0.5$, $0.4$ and $0.3$. The voxel's edge length
is $d=2.35\,$fm. The total number of nucleons is $6000$. (a)
$\rho=0.05\,$fm$^{-3}$ and (b) $\rho=0.085\,$fm$^{-3}$. }
\end{figure*}

Extending the $\chi$ study for non-symmetric nuclear matter appears to confirm 
the complexity observed in Sections~\ref{thermo} and \ref{kolmogorov_results}. 
The global pressure does not present negative values for $x=0.4$ and $x=0.3$ 
at temperatures above the solid-like state. Neither noticeable bubbles nor 
other 
heterogeneities could be detected at the early stage of the pasta formation
(say, $T\approx 2\,$MeV). We are now able to confirm these results through the
$\chi$ functional. Fig.~\ref{fig:mink_func} shows the Euler functional for
two different densities and $x$= 0.3, 0.4 and 0.5. \\

\begin{figure*}[!htbp]
\centering
\subfloat[$x=0.4$\label{fig:mink_slice_3_04}]{
\includegraphics[width=0.6\columnwidth]
{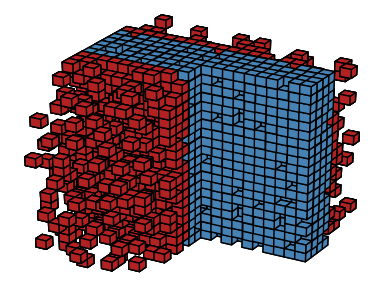}
} 
\hspace{3cm}\subfloat[$x=0.3$\label{fig:mink_slice_3_03}]{
\includegraphics[width=0.6\columnwidth]
{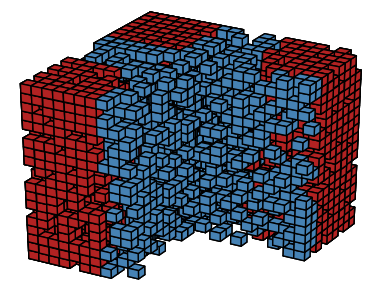}
} 
\caption{\label{fig:mink_slice_3} (Color online) Inside view of
discretized nuclear matter with 6000 nucleons
at $T=1.0\,$MeV, and $\rho=0.05\,$fm$^{-3}$. A quarter slice has been cut out
for a better view of the inner most region of the cell. The discretization
turned the 6000 nucleons into (approximately) 3800 voxels (edge length
$d=2.35\,$fm). (a)  Non-symmetric nuclear matter for $x=0.4$. The red color
corresponds to the region mostly occupied by neutrons (compare with
Fig.~\ref{fig:pasta_decomposed_3}). (b) Non-symmetric nuclear matter for
$x=0.3$. The blue color corresponds to the region mostly occupied by neutrons
(compare with Fig.~\ref{fig:pasta_decomposed_4}). }
\end{figure*}

Fig.~\ref{fig:mink_func_a} shows three distinct behaviors of 
$\chi$ at $\rho=0.05\,$fm$^{-3}$. The case of symmetric nuclear matter 
($x=0.5$)  was already analyzed above. The curve for $x=0.4$ appears 
left-shifted with respect to the symmetric case, in agreement with our previous
observation that proton ratios of $x<0.5$ frustrate for a while the pasta
formation (see Section~\ref{kolmogorov_results}). In spite of that, the pattern
for $x=0.4$ achieves a higher positive value at lower temperatures than the
symmetric case indicating that the tunnel-like scenario ($\chi<0$) switched to
a bubble-like or an isolated-structure scenario ($\chi>0$). It can be verified
from Fig.~\ref{fig:mink_slice_3_04} that this is actually occurring at
$T\approx 1\,$MeV; many isolated structures may be visualized in red, while no
tunnels seem to be present in the blue region (see caption for details). \\

The curve of $x=$ 0.3 in Fig.~\ref{fig:mink_func_a} does not 
change sign nor increases in value at lower temperatures, as opposed to the 
other two curves. The fact that $\chi < 0$ for all of the examined temperatures 
indicates that tunnel-like landscapes are the relevant ones. 
Fig.~\ref{fig:mink_slice_3_03} 
illustrates this scenario: the heavy tunnel-like region (highlighted in blue 
color) is mostly 
occupied by neutrons, as shown in Fig.~\ref{fig:pasta_decomposed_4}, 
indicating that repulsive forces between neutrons dominate in a large fraction 
of 
the cell, and thus producing a positive global pressure all along the examined 
temperature range, as noticed in Fig.~\ref{fig:energy_pressure}. 
Fig.~\ref{fig:mink_func_a} 
also show that $\chi$ for $x=0.3$ decreases in magnitude with at lower 
temperatures; 
this corresponds to the departure from the spacial homogeneity detected in 
Fig.~\ref{fig:rho05_pn}, and found by the Kolmogorov statistic and Euler 
functional 
to occur at $T\approx 1.5\,$MeV. \\

At the higher density of $\rho=0.085\,$fm$^{-3}$ the behavior of $\chi$ is 
substantially different. Fig.~\ref{fig:mink_func_b} shows the temperature 
dependence 
of $\chi$ for $x=$ 0.3, 0.4 and 0.5; both isospin asymmetric cases are 
qualitatively 
similar to the symmetric one. An inspection of the voxels' configuration (not 
shown) 
confirms that tunnels become relevant at very low temperatures; this can be 
checked 
from the configurations presented in Figs.~\ref{fig:pasta_decomposed} and 
\ref{fig:pasta_decomposed_2}.
\\

For obvious reasons the analysis of the $\chi$ functional cannot be applied to 
protons and neutrons separately, as it was done with the Kolmogorov statistic; 
excluding either protons or neutrons would produce fake voids, overestimating
the total number of tunnels or cavities. \\

We summarize the results from this section as follows. At $T\approx 2\,$MeV 
the Euler functional $\chi$ of isospin-symmetric low-density systems ($\rho< 
0.06\,$fm$^{-3}$) 
show drastic changes from negative to positive values indicating a transition 
from a void-dominated regime to one with bubbles and isolated regions. Higher 
density systems ($\rho>0.06\,$fm$^{-3}$), in spite of always having $\chi > 0$, 
also increase the value of their $\chi$ at this temperature, reaching a common 
maximum for all densities at $T\approx 1.5\,$MeV. This maximum corresponds 
to the formation of bubbles or isolated regions, and indicates the formation of 
the 
pasta near the solid-liquid transition; recall that the Kolmogorov statistic was 
able 
to detect the pasta formation since the bubbles or isolated regions stage. \\

For isospin asymmetric systems the low-density ($\rho\approx 0.05\,$fm$^{-3}$) 
growth of $\chi$ is also observed but only for $x=$ 0.4 and 0.5; systems at $x=$ 
0.3 
have $\chi<0$ at all temperatures. At higher densities ($\rho\approx 
0.08\,$fm$^{-3}$) 
the Euler functional is always positive for all temperatures. \\

Table~\ref{tab1} classifies the pasta according to the sign of $\chi$ and the 
curvature $B$, and it was a goal of the present study to extend this 
classification 
for isospin asymmetric systems, but our results indicate that this labeling 
becomes 
meaningless for the non-symmetric case. For a given temperature, the $\chi$ 
functional attains positive or negative values depending on the isospin content 
and 
the density of the system. In general, the excess neutrons obscure the pasta 
structures for the protons and, thus, the early stage of the pasta formation 
(that is, the formation of bubbles or isolated regions) is not detectable. In 
spite of this, we observe that the system departs from homogeneity 
at $T\sim 1.5\,$MeV (see, for example, Fig.~\ref{fig:rho05_pn}).  \\

\subsection{\label{subsec:esym}Symmetry energy and nuclear pasta}

We now study the symmetry energy of nuclear matter in the pasta region. As 
stated at the end of Section~\ref{cmd_star}, at a given temperature the energy 
$E(\rho)$ showed three distinct behaviors as a function of the density: the 
pasta region for densities below $0.085\,$fm$^{-3}$, the crystal-like region 
for 
densities above $0.14\,$fm$^{-3}$, and an intermediate region in between the 
first two. In what follows we will focus on the symmetry energy in the pasta 
region. \\

\begin{figure}
\begin{center}
   \includegraphics[width=\columnwidth]{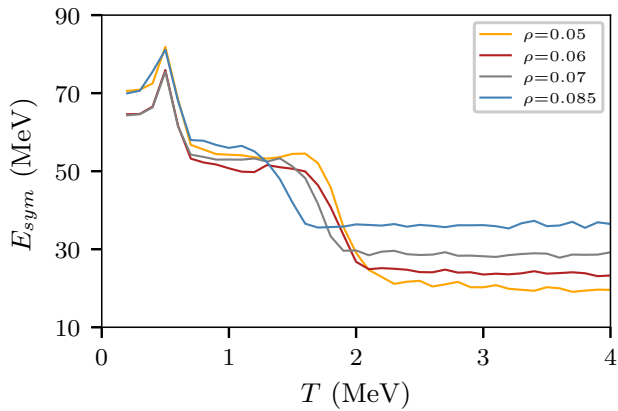}
\caption{(Color online) Symmetry energy as a function of the temperature
for four density values, as indicated in the insert. The system corresponds to
nuclear matter with 6000 nucleons.
}\label{fig:esym}
\end{center}
\end{figure}

Fig.~\ref{fig:esym} shows the computed symmetry energy as a function of the 
temperature for the four examined densities. The $E_{sym}$ was computed through 
the fitting procedure outlined in Section~\ref{esymm}. An analysis of the 
fitting errors can be found in Appendix~\ref{fitting_errors}. \\

Several distinct regions can be distinguished for $E_{sym}$ in 
Fig.~\ref{fig:esym}.
In cooling, a liquid system with $T>2\,$MeV starts with a low value of 
$E_{sym}$. 
Upon entering the $T<2\,$MeV region and until $T\approx 1.5$ MeV, the symmetry 
energy increases in magnitude while the liquid pasta is formed.  Its value 
stabilizes 
in an intermediate value in the warm-to-low temperature range of $0.5$ 
MeV$<T<1.5$ 
MeV where the liquid pasta exists. At $T<0.5$ MeV, when the liquid-to-solid 
phase 
transition happens within the pasta, the symmetry energy reaches its highest 
value. 
We now look at these different stages in turn. \\

In the higher temperature range, $T>2\,$MeV, the symmetry energy has different 
magnitudes for the four densities explored (see  Fig.~\ref{fig:esym}), with the 
higher values of $E_{sym}$ corresponding to higher densities. Noticing that 
this 
relationship is also maintained by the Euler functional ($\chi$) shown in 
Fig.~\ref{mink_x05}, it is possible to think that there is a connection between 
the symmetry energy and the morphology of the system. Remembering from 
Section~\ref{minkowski_results} that at those temperatures higher densities 
are associated to cavity-like or isolated regions, and lowest densities with 
tunnel-like structures, it is safe to think that $E_{sym}$ increases as the 
tunnels 
become obstructed and cavities or isolated regions prevail.   \\ 

\begin{figure*}[!htbp]
\centering
\subfloat[$\rho=0.05$]{
\includegraphics[width=\columnwidth]
{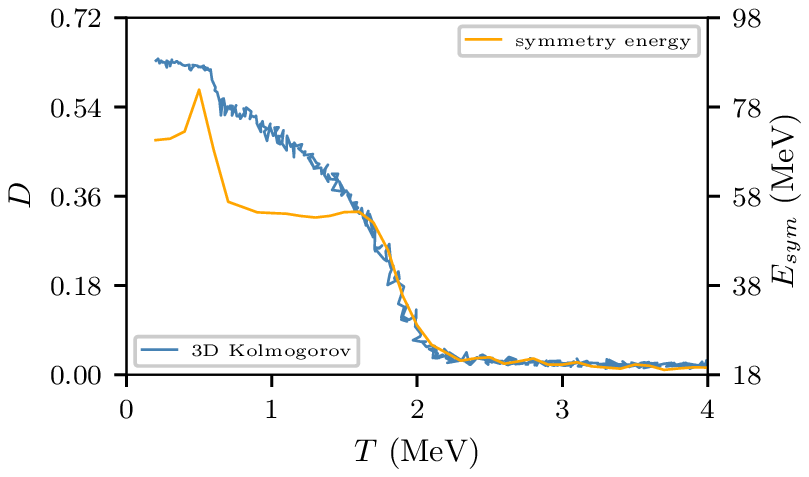}
} 
\subfloat[$\rho=0.06$]{
\includegraphics[width=\columnwidth]
{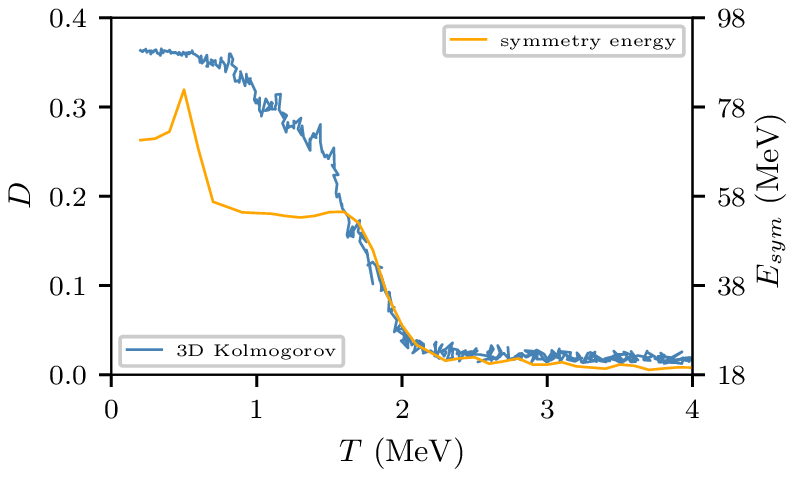}
} 
\\
\subfloat[$\rho=0.07$]{
\includegraphics[width=\columnwidth]
{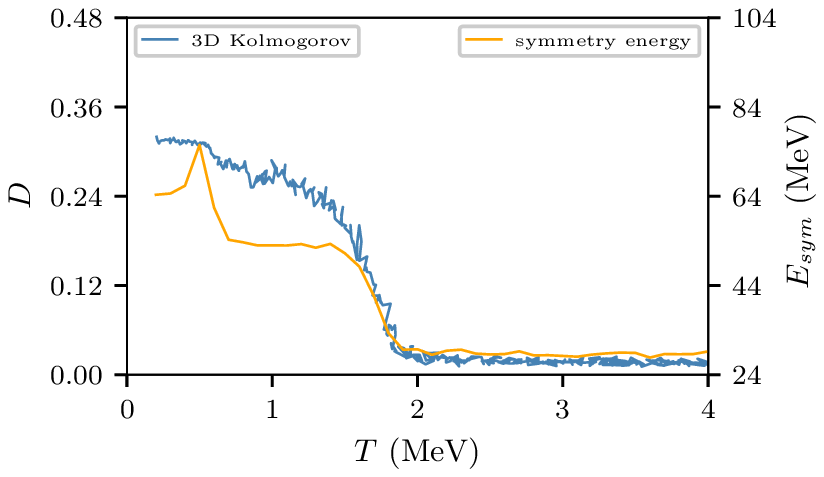}
} 
\subfloat[$\rho=0.085$]{
\includegraphics[width=\columnwidth]
{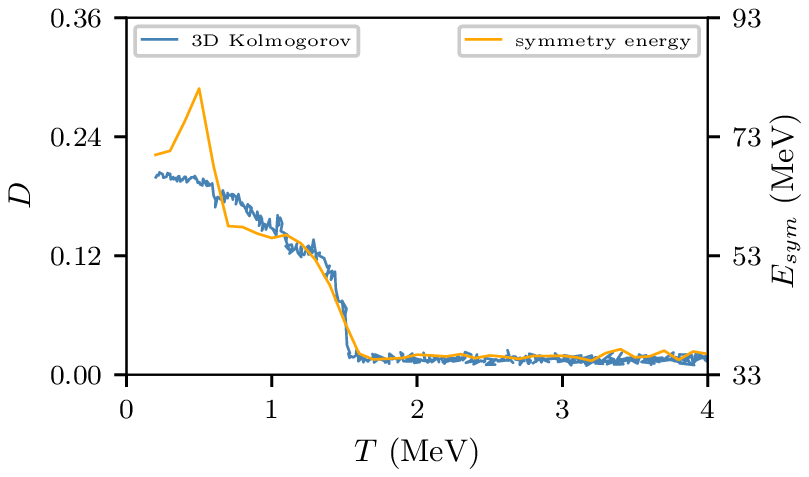}
} 
\caption{\label{fig:esym_kolmogorov} (Color online) Symmetry energy
(right scale) and 3D Kolmogorov statistic (left scale) as a function of the
temperature, for nuclear matter with 6000
nucleons at $x=0.5$. The corresponding densities are indicated below each plot
(units fm$^{-3}$). }
\end{figure*}

Between $1.5$ MeV $<T<2$ MeV the symmetry energy appears to vary 
in a way resembling the variation of the 3D Kolmogorov statistic, $D$, during 
the 
pasta formation stage. To test this hypothesis Fig.~\ref{fig:esym_kolmogorov} 
compares the variation of $E_{sym}$ with that of $D$ as a function of 
temperature 
for the four densities of interest, and finds a good match between both 
quantities. 
This interesting observation allows us to use the reasonings in 
Section~\ref{minkowski_results} to conclude that the variation of $E_{sym}$ can 
be associated to the changes of the Euler functional $\chi$ (see 
Fig.~\ref{mink_x05}), which in turn are associated to the morphology of the 
nuclear matter structure. Indeed one can conclude that as the pasta is formed 
during cooling, the symmetry energy increases in magnitude. \\

The value of $E_{sym}$ between $0.5$ MeV $<T<1.5$ MeV corresponds to 
the pasta structures filled with liquid nuclear matter. Its change at around 
$T\approx 0.5$ MeV happens at the same temperature at which the caloric 
curve and the Lindemann coefficient undergoes similar changes (cf. 
Fig.~\ref{lin}), 
indicating the phase transition between a liquid pasta ($T>$ 0.5 MeV) and a 
solid 
pasta ($T<$ 0.5 MeV). Here one can concluded that the symmetry energy 
attains its largest value in solid pastas. \\

Finally, it is worth noting that the values of $E_{sym}$ attained in the 
studied 
densities and temperatures are not directly comparable to those calculated 
in~\cite{lopez2017} for the liquid-gas coexistence region and 
compared to experimental results. This lack of comparability comes, first, 
because the calculation of Fig.~4 in~\cite{lopez2017} was obtained at higher 
temperatures ($2$ MeV $< T < 5$ MeV) and lower densities 
($\rho<$ 0.03$\,$fm$^{-3}$), and, second, such calculation was performed 
with data from a homogeneous medium while the present one uses data 
from a pasta-structured system. In spite of these differences, it is reassuring 
that the values of $E_{sym}$ obtained in the present study for the highest 
temperatures used in the present study, $T>2$ MeV, are within the range 
of values calculated in~\cite{lopez2017} for the highest densities considered 
in such study, namely 0.05 fm${}^{-3}<\rho<$ 0.06 fm${}^{-3}$.
\\

\section{Conclusions}\label{concluding}

In this article we have investigated the formation of the pasta in nuclear 
matter (according to the CMD model), its phase transitions and its symmetry 
energy. In particular we focused in isospin symmetric ($x=0.5$) and isospin 
asymmetric ($x=0.3$ and 0.4) nuclear matter. We explored, first, if the pasta 
structures found in symmetric nuclear matter exist in non-symmetric systems 
and, 
second, if the solid-to-liquid phase 
transitions found in Ref.~\cite{dorso2014} survived in neutron-rich nuclear 
matter. Lastly, we studied the behavior of the symmetry energy in systems with 
pasta configurations and whether it is connected to the isospin content, the 
morphology of the pasta and to the phase transitions.\\

After introducing a plethora of simulation and computational tools, molecular 
dynamics simulations were used to produce homogeneous systems that, when cooled 
down below $T\approx2$ MeV, were reconfigured into pasta structures; this 
happened for systems with proton to neutron ratios of $x=0.3$, 0.4 and 0.5, 
densities in the range of 0.05 fm${}^{-3}\rho<$ 0.08 fm${}^{-3}$, and were 
observed by directly drawing the structures, 
cf. Fig.~\ref{fig:pasta}.\\

Studying the caloric curve (cf. Figs.~\ref{cc} and~\ref{fig:energy_pressure}),  
and the Lindemann coefficient (cf. Fig.~\ref{lin}) allowed the detection of 
phase transitions as the nuclear matter system cooled down. The slopes of the 
caloric curves and of the pressure-temperature curves were also found to vary 
with density as seen in Fig.~\ref{fig:energy_pressure}. A metastable region 
(i.e. negative pressure) appears to exist only for symmetric systems but not 
for 
neutron-rich media.\\

The radial distribution function allowed the study of the inner structure of 
the 
matter filling the pasta structures. Figs.~\ref{rad} and~\ref{rad2} indicate 
that at very low temperatures ($T<0.5\,$MeV) the pasta is in a solid state, 
but at higher temperatures ($0.5\,$ MeV $< T<1.5\,$MeV) it transitions to 
a liquid phase while maintaining the pasta structure. \\

Continuing with a study of the departure from homogeneity, the 1D Kolmogorov 
statistic $D$ was applied separately on each Cartesian coordinate. 
Figure~\ref{fig:rho05_x05_t20_all} showed homogeneity in isospin symmetric and 
non-symmetric systems at temperatures above $2\,$MeV, but at $T\approx 2\,$MeV, 
$D_y$ showed a definite departure from homogeneity for $T<2\,$MeV in agreement 
with the formation of lasagnas seen in Fig.~\ref{fig:rho05}. Applying the 1D 
Kolmogorov statistic on protons and neutrons separately shows that in 
non-symmetric systems the signal from the 1D Kolmogorov statistic appears 
stronger for protons than for neutrons indicating that in neutron-rich systems 
the excess neutrons delay the formation of the pasta formation while the 
protons 
indeed form pasta structures at lower temperatures. \\

This complex behavior was corroborated by the Euler functional $\chi$ which 
showed jumps at the same temperatures at which the phase changed. $\chi$ was 
found to be correlated, depending on the density, with the existence of 
tunnels, 
empty regions, cavities and isolated regions. Indeed, at $T\approx 2\,$MeV in 
isospin-symmetric low-density systems ($\rho< 0.06\,$fm$^{-3}$) changes of 
$\chi$ reflect transitions from a void-dominated to a bubble-dominated region, 
and at higher densities ($\rho>0.06\,$fm$^{-3}$) $\chi$ reaches a maximum 
when the pasta forms at the solid-liquid transition. Likewise, for isospin 
asymmetric systems at low-densities ($\rho\approx 0.05\,$fm$^{-3}$) $\chi$ 
attained positive or negative values depending on the isospin content and the 
density of the system. \\

All of these phase changes that occur during cooling were reflected in the 
values of the symmetry energy, which showed different behavior as a function of 
the temperature (cf. Fig.~\ref{fig:esym}). At the higher temperatures 
($T>2\,$MeV), $E_{sym}$ had different magnitudes for each of the four 
densities studied, indicating that in homogeneous systems the symmetry energy 
increases with the density.  In the transition from homogeneity to a 
liquid-filled pasta $E_{sym}$ varies in the same way as the 3D Kolmogorov 
statistic $D$, and the Euler functional $\chi$, showing its dependence on the 
morphology of the nuclear matter structure. The transition from a liquid pasta 
($T>$ 0.5 MeV) to a solid pasta ($T<$ 0.5 MeV) was also reflected on the value 
of $E_{sym}$ with an increase to its largest value. \\

In conclusion, classical molecular dynamics simulations show the formation of 
pastas in isospin symmetric and non-symmetric systems. The computational tools 
developed and applied, although not perfect, demonstrated their usefulness to 
detect the in-pasta phase transitions first seen in Ref.~\cite{dorso2014}, 
and to extend the calculation of the symmetry energy of Ref.~\cite{lopez2017} 
to 
lower temperatures, and connect its value to the structure and thermodynamics 
of 
the neutron-rich pasta.  \\

We acknowledge that, although useful for a first study, the used tools produced 
a somewhat fuzzy picture of the nucleon dynamics within the pasta, and that  
these tools need to be refined in the future. In summary the pasta structures 
were fairly detected at temperatures as high as $T\sim 2\,$MeV, the radial 
distribution function gave information to infer a phase transition within the 
pasta, and the Kolmogorov statistic and the Minkowski functional were 
useful in pointing at the early stage of the pasta formation. Future studies 
will try to refine these tools and apply them to study the formation of the 
pasta, possible phase transitions and the role of the symmetry energy in 
neutron-rich neutron star matter. \\

\begin{acknowledgments}
Part of this study was financed by FONCyT (Fondo para la
Investigaci\'on Cient\'\i fica y Tecnol\'ogica) and Inter-American Development
Bank (IDB),  Grant Number PICT 1692 (2013).
\end{acknowledgments}

\appendix

\section{The radial distribution $g(r)$}\label{sec:gr_example}

A rigorous definition for the radial distribution function $g(r)$ starts from
the following distance distribution

\begin{equation}
g(\mathbf{r})=\displaystyle\frac{1}{\rho_0}\bigg\langle\displaystyle\frac{1}{N}
\displaystyle\sum_{i=1}^N\displaystyle\sum_{j\neq i}^N
\delta(\mathbf{r}-\mathbf{r}_{ij} )\bigg\rangle\label{eqn:gr_def_rigorous}
\end{equation}

\noindent where $\rho_0=N/V$ is the (mean) density in the simulation cell of
volume $V$ (or equivalently $L^3$).
$\mathbf{r}_{ij}=\mathbf{r}_{j}-\mathbf{r}_{i}$ is the distance
vector between the particle $i$ and the particle $j$. The $\delta(\cdot)$
function corresponds to the Dirac
delta. The mean value indicated by $\langle\cdot\rangle$ corresponds to the
average operation over successive time-steps. \\

In order to illustrate the pattern of the $g(r)$ due to a pasta
structure, we assume that particles form a homogeneous slab (that is, a
\textit{lasagna}-like structure) bounded between the $\pm z_0$ (horizontal)
planes. The slab is quasi static, meaning that no averaging over successive
time-steps is required. \\

We first consider a small region $\Omega(\mathbf{r})$ with volume $\delta
x\,\delta y\,\delta z$ in the $\mathbf{r}$ domain. Thus, we may evaluate the
expression (\ref{eqn:gr_def_rigorous}) as follows

\begin{equation}
\displaystyle\int_{\Omega(\mathbf{r})}g(\mathbf{r})\,d^3x=
\displaystyle\frac{1}{\rho_0}\,
\displaystyle\frac{1}{N}
\displaystyle\sum_{i=1}^N \sum_{j\neq i}^N \int_{\Omega(\mathbf{r})}
\displaystyle\delta(\mathbf{r}-\mathbf{r}_{ij})\,d^3x
\label{eqn:gr_def_rigorous_2}
\end{equation}

\noindent where the integral on the right-hand side equals unity at the
positions $\mathbf{r}=\mathbf{r}_{ij}$ within $\Omega(\mathbf{r})$. \\

We now proceed to evaluate the sum over the $j\neq i$ neighbors of the
particle $i$. The counting of neighbors is proportional to $\delta x\,\delta
y\,\delta z$ since the volume $\Omega(\mathbf{r})$ is very small. Thus, the
tally is

\begin{equation}
\sum_{j\neq i}^N \int_{\Omega(\mathbf{r})}
\displaystyle\delta(\mathbf{r}+\mathbf{r}_{i}-\mathbf{r}_{j})\,d^3x\approx
\rho(\mathbf{r}+\mathbf{r}_i)\,\delta x\,\delta y\,\delta z
\label{eqn:gr_def_rigorous_3}
\end{equation}

The magnitude $\rho(\mathbf{r}+\mathbf{r}_i)$ refers to the density within the
small region $\Omega(\mathbf{r}+\mathbf{r}_i)$, its value remains constant
inside the slab and vanishes outside. In fact, $\rho$ can be expressed as
$\rho_s\,\Theta(z_0-|z+z_i|)$, with $\Theta(\cdot)$ representing the Heaviside
function, and $\rho_s$ being the slab density ($\rho_s>\rho_0$).   \\

The sum over the $i$ particles is evaluated through the integral of the 
infinitesimal 
pieces $\rho_s\,d^3x'$. Thus,

\begin{equation}
\displaystyle\sum_{i=1}^N
\rho(\mathbf{r}+\mathbf{r}_i)\approx\displaystyle\int_\mathrm{slab}\!\!\rho_s^2
\,\Theta(z_0-|z+z'|)\,d^3x'
\label{eqn:gr_def_rigorous_4b}
\end{equation}

\noindent where the slab domain corresponds to the bounded region $|z'|\leq
z_0$. Notice that this integral vanishes for $|z|\geq 2z_0$ and equals
$\rho_s^2L^2\,(2z_0-|z|)$ otherwise. The $g(\mathbf{r})$ function is
then

\begin{equation}
g(\mathbf{r})\approx \displaystyle\frac{\rho_s}{\rho_0}\,\bigg(1-
\displaystyle\frac{|z|}{2z_0}\bigg)\ \ \ , \ \ \ |z|<2z_0
\label{eqn:gr_def_rigorous_4}
\end{equation}

\noindent assuming that $g(\mathbf{r})$ is somewhat fixed inside the small
domain $\Omega(\mathbf{r})$ in (\ref{eqn:gr_def_rigorous_2}). We further
replaced $N$ by $\rho_s\,2z_0L^2$ in (\ref{eqn:gr_def_rigorous_2}). \\

The expression (\ref{eqn:gr_def_rigorous_4}) is not exactly the
radial distribution function yet because of the angular dependency of
$g(\mathbf{r})$. This dependency may be eliminated by integrating 
$g(\mathbf{r})$
over a spherical shell of radius $r$, which introduced the normalization factor 
$4\pi r^2$ in agreement with Eq.~(\ref{eqn:gr_def}). The expression
for the radial distribution then reads

\begin{equation}
g(r)=\displaystyle\frac{1}{4\pi r^2}\,
\displaystyle\int_S \displaystyle\frac{\rho_s}{\rho_0} \, \bigg(1-
\displaystyle\frac{r|\cos\theta|}{2z_0}\bigg)\,r^2\sin\theta\,d\theta\,d\varphi
\label{eqn:gr_def_rigorous_5}
\end{equation}

\noindent where $z=r\,\cos\theta$. Notice that this expression is valid along
the interval $|\cos\theta|<2z_0/r$ whenever $2z_0<r$, but it is constrained
to the natural bound $|\cos\theta|\leq 1$ if $2z_0\geq r$. The integration 
of (\ref{eqn:gr_def_rigorous_5}) finally yields

\begin{equation}
g(r)=\left\{\begin{array}{lcl}
             \displaystyle\frac{\rho_s}{\rho_0} \,
\bigg(1-\displaystyle\frac{r}{4z_0}\bigg) & \mathrm{if} & r<2z_0\\
             & & \\
             \displaystyle\frac{\rho_s}{\rho_0}\,\displaystyle\frac{z_0}{r}  &
\mathrm{if}  &  r>2z_0 \\
            \end{array}\right.\label{eqn:gr_def_rigorous_6}
\end{equation}

Notice that as $r\rightarrow0$, $g(r)$ correctly goes to the limit of 
$g(r) \rightarrow \rho_s/\rho_0$. Similarly, for larger values of $r$, and up 
to $r<2z_0$, $g(r)$ decreases linearly, as observed in Fig.~\ref{rad2} 
for different values of $\rho_0$. \\

Eq. (\ref{eqn:gr_def_rigorous_6}) also indicates that $g(2z_0)\approx 1$ in the 
case
that $\rho_s/\rho_0\approx 2$, in agreement with Fig.~\ref{rad2} where $g(r)$ 
tends to unity at $20\,$fm (for simulation cells of $L\sim 45\,$fm); this 
figure, 
however, does not show the behavior beyond $20\,$fm as the statistics 
that can be collected for distances above $L/2$ are very poor. \\

From literature references, $g(r)$ is supposed to converge to unity as
$r\rightarrow\infty$. But, according to (\ref{eqn:gr_def_rigorous_6}), the
radial distribution vanishes at this limit. The disagreement corresponds to
the fact that the condition $g(\infty)=1$ is only valid for homogeneous
systems. The expression (\ref{eqn:gr_def_rigorous_6}) can actually meet this
condition if the slab occupies all the simulation cell, since
$\rho_s\rightarrow\rho_0$ and $z_0\rightarrow\infty$ (for periodic boundary
conditions). \\

\section{\label{sec:voxel}The Minkowski voxels}

The Minkowski functionals require the binning of space into ``voxels''
(that is, tridimensional ``pixels''). Each voxel is supposed to include
(approximately) a single nucleon. But this is somehow difficult to achieve if
the system is not completely homogeneous (and regular). \\

\begin{figure*}[!htbp]
\centering
\subfloat[]{
\includegraphics[width=\columnwidth]
{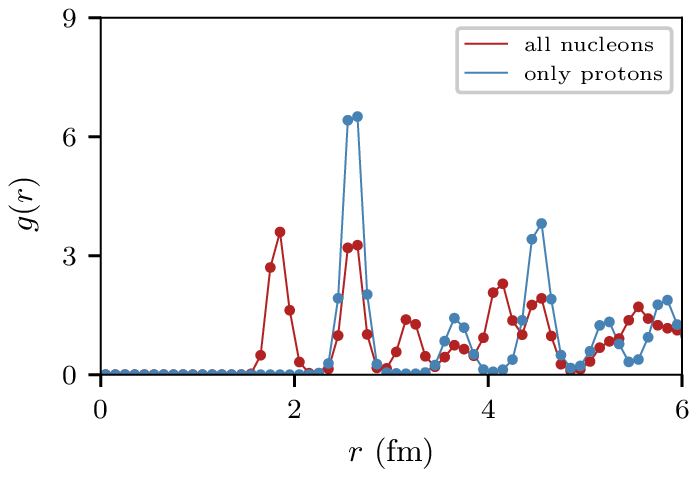}
} 
\hspace{2.5cm}\subfloat[]{
\includegraphics[width=0.55\columnwidth]
{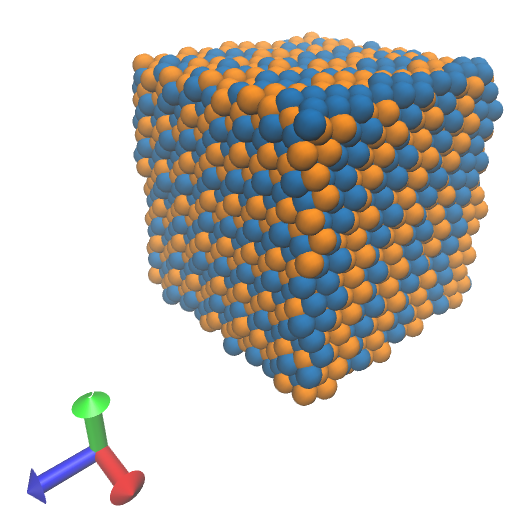}
}
\caption{\label{fig:rho016_x05} (Color online) (a) Radial
distribution function $g(r)$ for the case of 5832 nucleons ($x=0.5$) at
$\rho=0.16\,$fm$^{-3}$ and $T=0.1\,$MeV. The binning is $0.05\,$fm. The symbols
in blue correspond to the $g(r)$ for protons only. The symbols in red correspond
to the $g(r)$ computed over all the nucleons. The first peak for the blue
symbols occurs at $2.65\,$fm. The first peak for the red symbols occurs at
$1.85\,$fm. (b) A snapshot of the system at $T=0.1\,$MeV.  }
\end{figure*}

We start with a simple cubic arrangement of 50\% protons and 50\% neutrons as
shown in Fig.~\ref{fig:rho016_x05}. The system is at the saturation density
$\rho_0=0.16\,$fm$^{-3}$ (see caption for details).  \\

Notice from Fig.~\ref{fig:rho016_x05} that the nearest distance between protons
and neutrons is $1.85\,$fm. Likewise, the nearest distance between nucleons of
the same species is approximately $2.65\,$fm (that is, the position of the
second peak). It can be verified that the latter is approximately
$\sqrt{2}\times 1.85\,$fm, as expected for the simple cubic arrangement (within
the
$g(r)$ binning errors).  \\

\begin{figure*}[!htbp]
\centering
\subfloat[\label{fig:rho016_x05_chi_euler}]{
\includegraphics[width=\columnwidth]
{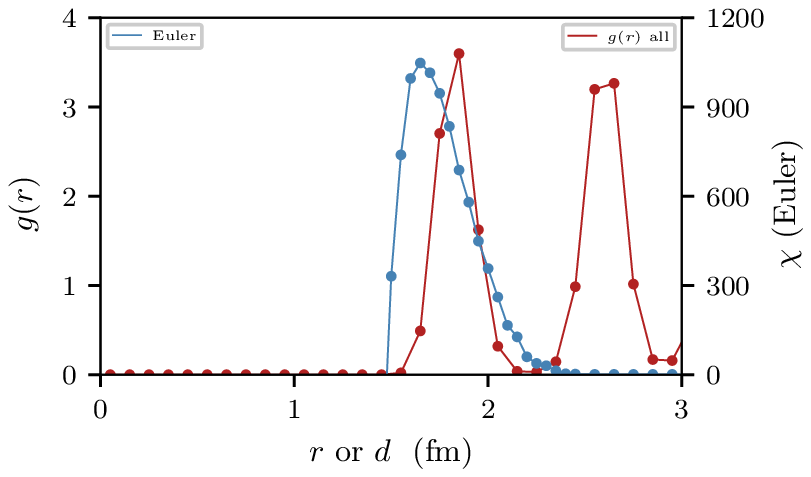}
} 
\subfloat[\label{fig:rho016_x05_chi_vol}]{
\includegraphics[width=\columnwidth]
{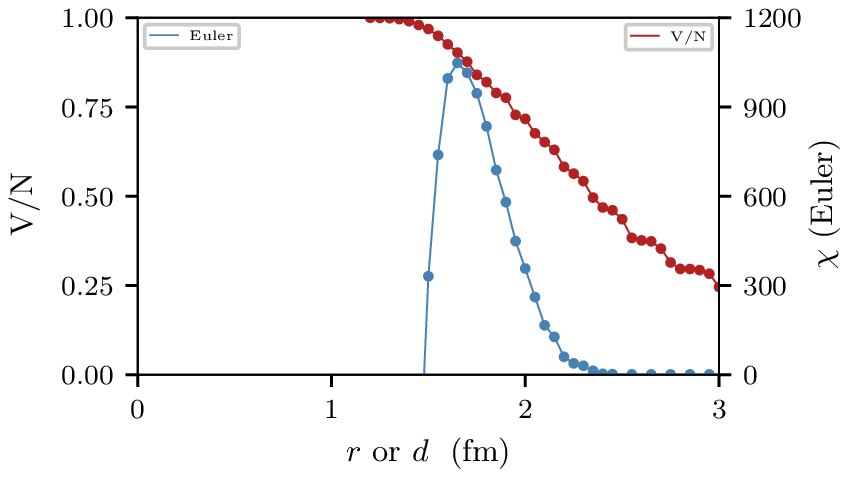}
}
\caption{\label{fig:rho016_x05_chi} (Color online) Analysis of the
same system as in Fig.~\ref{fig:rho016_x05} (with $N=5832$ nucleons). (a) On
the left scale (red symbols), the radial distribution function $g(r)$. On the
right scale, the Euler functional $\chi$ as a function of the voxel's edge
length $d$. (b) On the left scale (blue symbols), the first Minkowski
functional (volume) as a function of the voxel's edge length $d$. The volume is
normalized by $N$. On the right scale, the same the Euler functional $\chi$ as
in (a) for comparison reasons. }
\end{figure*}

Fig.~\ref{fig:rho016_x05_chi_euler} reproduces the same pattern for the
distribution $g(r)$ over all the nucleons. It further shows the Euler
functional $\chi$ as a function of the voxel's width (see caption for details).
Both curves share the same abscissa for comparison reasons. For small values of
$d$ the functional $\chi$ is negative (not shown), meaning that the voxels are
so small that tunnels prevail in the (discretized) system. At $d=1.65\,$fm this
functional arrives to a maximum where cavities or isolated regions prevail.
Some cavities may be ``true'' empty regions but other may simply be fake voids.
The most probable ones, however, correspond to fake voids since the $g(r)$
pattern actually presents a maximum at $r>1.65\,$fm. Thus, increasing the
voxel's size will most probably cancel the fake cavities.  \\

The particles located at the first maximum of $g(r)$ (at the saturation
density) may be envisaged as touching each other in a regular (simple cubic)
array. Therefore, the mean radius for a nucleon should be $1.85/2\,$fm. This
means, as a first thought, that binning the space into voxels of width
$1.85\,$fm will include a single nucleon per voxels. This is, however, not
completely true since approximately half of the first neighbors exceeds the
$1.85\,$fm (see first peak in Fig.~\ref{fig:rho016_x05}). Many voxels will be
empty, and thus, a relevant probability of finding fake voids exists.
Fig.~\ref{fig:rho016_x05_chi_euler} illustrates this situation. \\

\begin{figure*}[!htbp]
\centering
\subfloat[$d=1.85$\label{fig:voxels_width_a}]{
\includegraphics[width=0.85\columnwidth]
{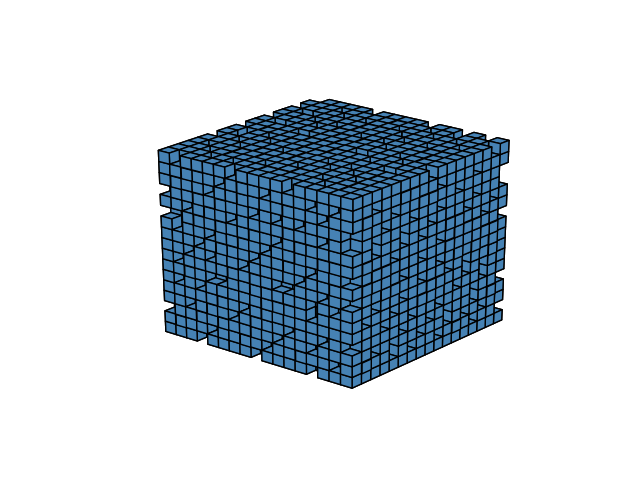}
} 
\subfloat[$d=2.35$\label{fig:voxels_width_b}]{
\includegraphics[width=0.85\columnwidth]
{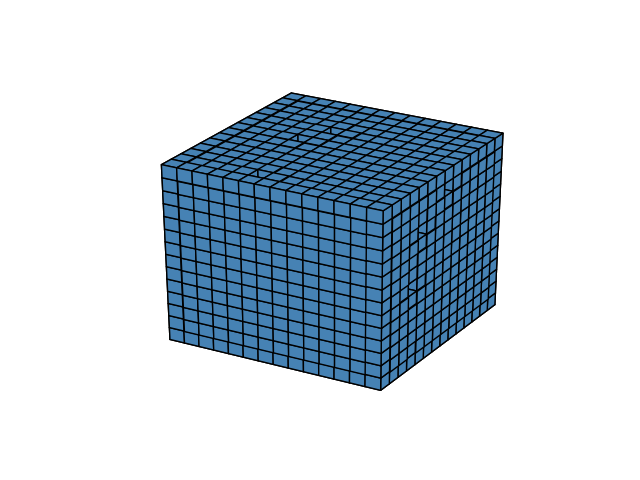}
} 

\caption{\label{fig:voxels_width} Example of the space binning into voxels
corresponding to the situation shown in Fig.~\ref{fig:rho016_x05}. (a) The edge
length of the cubic voxel is $d=1.85\,$fm. (b) The edge length of the cubic
voxel is $d=2.35\,$fm. }
\end{figure*}

Notice that whenever a fake empty voxel exists, the contiguous one will perhaps
host two nucleons. This is because the nucleon that exceeds the $1.85\,$fm
distance to the neighbor, say on the left, may have shorten the distance to the
neighbor on the right. Thus, the number of occupied voxels will probably not
match the number of nucleons. Fig.~\ref{fig:rho016_x05_chi_vol} shows a decrease
in the number of occupied voxels (\textit{i.e.} the volume) for $r>1.6\,$fm. \\

The space binning should be done wider in order to to avoid fake empty voxels.
But, too wide voxels may include second neighbors. The most reasonable binning
distance appears to be around the first minimum of $g(r)$ (see
Fig.~\ref{fig:rho016_x05}). That is, at some point between $2.15\,$fm and
$2.35\,$fm. \\

A reasonable \textit{criterion} for the space binning may raise from the
Euler functional: the right binning distance should drive the $\chi$ functional
to unity, that is, to a single compact region. This occurs at $d=2.35\,$fm for
sure, as can be seen in Fig.~\ref{fig:voxels_width_b}. It can be further
checked from Fig.~\ref{fig:rho016_x05_chi_vol} that this binning allows the
hosting of approximately two nucleons per voxels, meaning that the
number of fake voids is negligible. \\

\section{\label{fitting_errors}The $E_{sym}$ fitting procedure errors}

\begin{figure*}[!htbp]
\centering
\subfloat[$\rho=0.05$]{
\includegraphics[width=\columnwidth]
{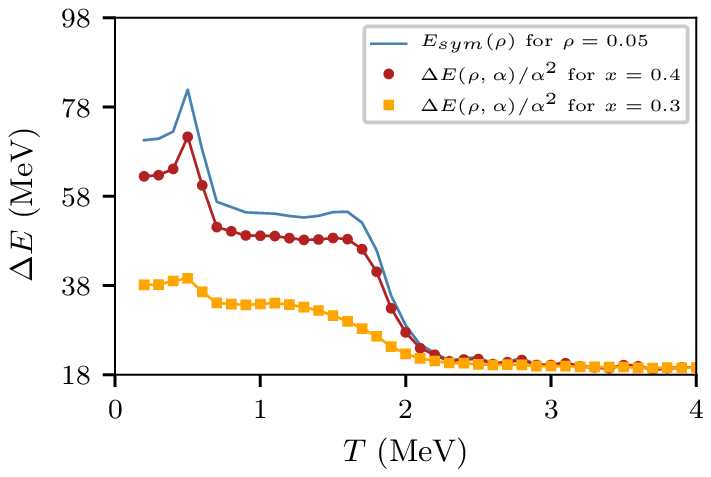}
} 
\subfloat[$\rho=0.06$]{
\includegraphics[width=\columnwidth]
{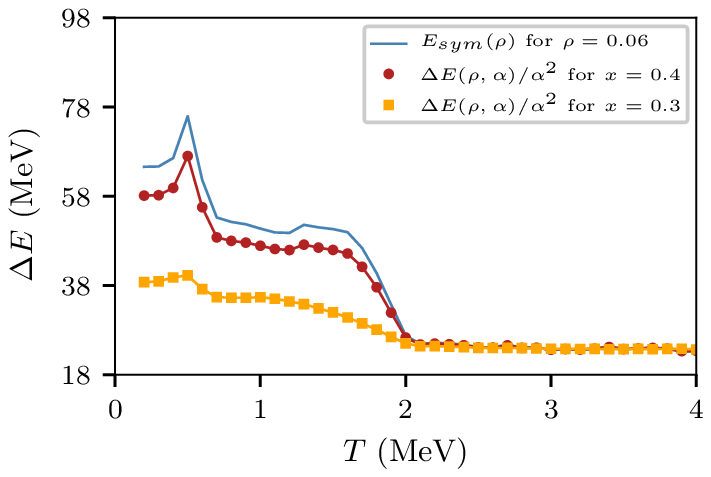}
} 
\\
\subfloat[$\rho=0.07$]{
\includegraphics[width=\columnwidth]
{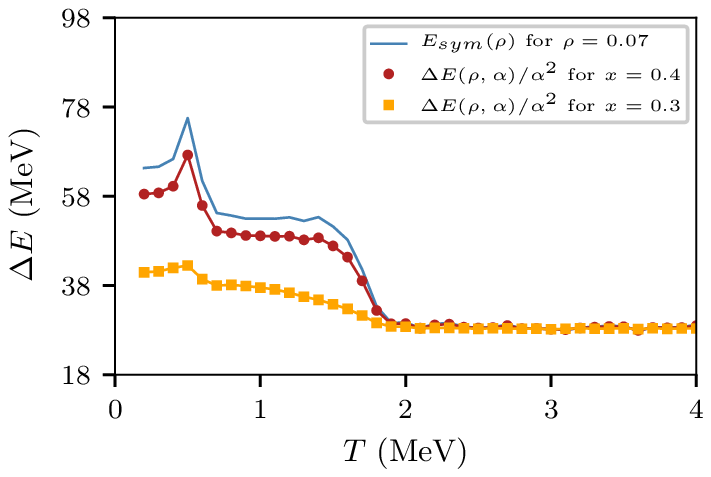}
} 
\subfloat[$\rho=0.085$]{
\includegraphics[width=\columnwidth]
{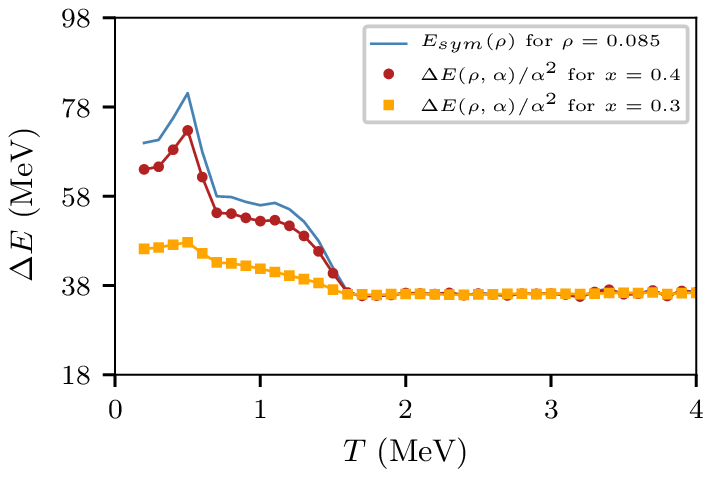}
} 
\caption{\label{fig:esym_test} Symmetry energy and the variation of
internal energy $\Delta E(\alpha)$ as a function of the temperature, for
nuclear
matter with 6000 nucleons. The corresponding
densities are indicated below each plot (units fm$^{-3}$). Rounded symbols
correspond to $\Delta E(\alpha=0.2)$ or $x=0.4$. Squared symbols correspond to
$\Delta E(\alpha=0.4)$ or $x=0.3$. The single line represents the fitting
procedure (see text for details).  }
\end{figure*}

The accuracy of the fitting procedure was further tested according to the
hypothetical expansion

\begin{equation}
\Delta E(\alpha)\simeq
E_{sym}\,\alpha^2+\mathcal{O}(\alpha^4)\label{eq:expansion}
\end{equation}

\noindent where $\Delta E=E(\rho,T,\alpha)-E(\rho,T,\alpha=0)$. The odd-terms
in
$\alpha$ are excluded due to the exchange symmetry between protons and neutrons
of the strong force \cite{lopez2014}. Fig.~\ref{fig:esym_test} shows the
corresponding results for the explored densities.\\

\begin{figure*}[!htbp]
\centering
\subfloat[$x=0.4$]{
\includegraphics[width=\columnwidth]
{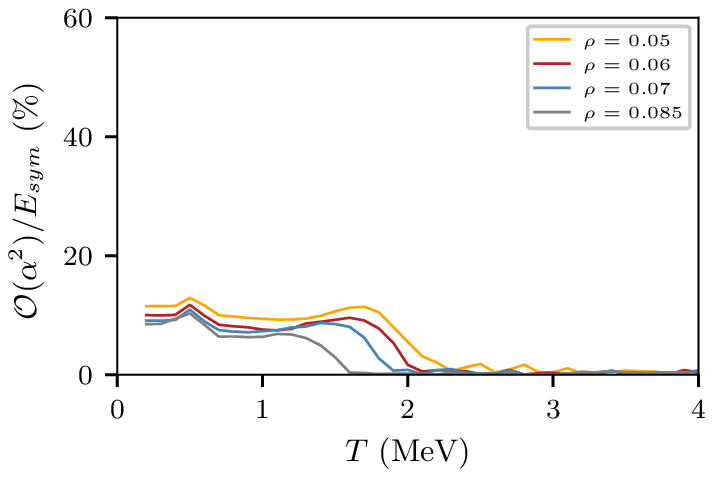}
} 
\subfloat[$x=0.3$]{
\includegraphics[width=\columnwidth]
{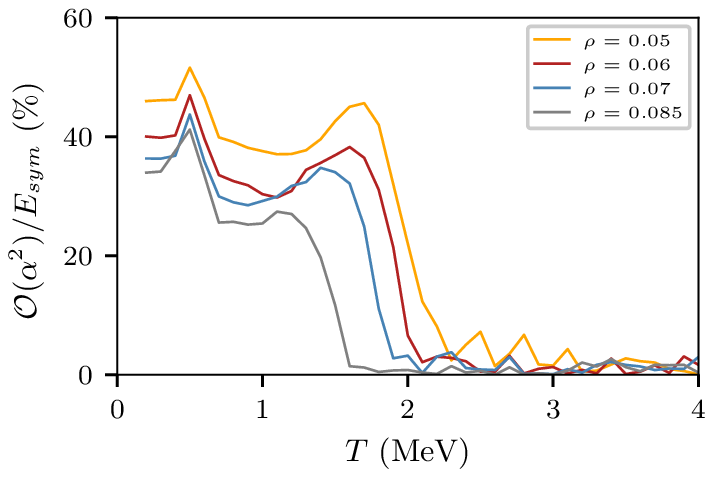}
} 
\caption{\label{fig:esym_test_2} (Color online) Relative difference for
the high order terms with respect to the computed symmetry energy (in modulus
and expressed in percentage) as a function of the temperature. The system
corresponds to nuclear matter with 6000
nucleons. On the left, the explores densities at $x=0.4$. On the right, the
explores densities at $x=0.3$. The corresponding densities are indicated in the
insert  (units fm$^{-3}$). }
\end{figure*}

The current fitting procedure appears to be accurate if compared to the
ratio $x=0.4$, but a noticeable bias is present for the ratio $x=0.3$. This
means that the $\mathcal{O}(\alpha^2)$  terms (that is,
$\mathcal{O}(\alpha^4)/\alpha^2$) may become relevant for proton fractions as
low as 30\%. Nevertheless, the fitting procedure always results in $\Delta
E<E_{sym}$ (despite round-off errors), meaning that the higher order
corrections
should carry a negative sign. Fig.~\ref{fig:esym_test_2} shows the relative
higher order discrepancy (in modulus) for $x=0.4$ and $x=0.3$, respectively.\\

The higher order terms, at the pasta temperature range, represent
roughly a 10\% correction with respect to the first order approach of $E_{sym}$
for $x=0.4$, and around 40\% for $x=0.3$, respectively. However, we checked over
that both corrections have essentially the same pattern, regardless of a
scaling factor. \\

Although the patterns shown in Fig.~\ref{fig:esym_test_2} and
Fig.~\ref{fig:esym} are very similar, the former exhibits a local maximum
within the temperature range $1-2\,$MeV. This maximum appears after the
computation of $E_{sym}-\Delta E(\alpha)/\alpha^2$. It can be seem from
Fig.~\ref{fig:esym_test} that the difference in the slopes between $E_{sym}$
and $\Delta E(\alpha)$ is responsible for this phenomenon. Thus, the
rate at which $E_{sym}$ jumps from the high temperature regime to the
pasta regime cannot be currently analyzed through the first order
approach in Eq.~(\ref{eq:expansion}), but through the higher order terms.  \\

\end{document}